\documentclass[reprint,nofootinbib,superscriptaddress,amsmath,amssymb,aps]{revtex4-1}
\usepackage{graphicx}
\usepackage{rotating}
\usepackage{dcolumn}
\usepackage{bm}
\usepackage{color}
\usepackage{mathptmx, textcomp}
\usepackage[utf8]{inputenc}
\usepackage{braket}
\usepackage[colorlinks,linkcolor=magenta,anchorcolor=cyan,citecolor=blue]{hyperref}
\usepackage[multiple]{footmisc}

\def\be{\begin{equation}}
\def\ee{\end{equation}}
\def\ba{\begin{eqnarray}}
\def\ea{\end{eqnarray}}

\newcommand{\eq}[1]{(\ref{#1})}

      \def\p {\pi} \def\a {\alpha}               \def\p {\pi}   
              \def\grad{\nabla}\def\.{\cdot}

\begin{document}

\title{Overspinning a rotating black hole in semiclassical gravity with type-A trace anomaly}
\author{Jie Jiang}
\email{jiejiang@mail.bnu.edu.cn}
\affiliation{College of Education for the Future, Beijing Normal University, Zhuhai 519087, China}
\author{Ming Zhang}
\email{mingzhang@jxnu.edu.cn (Corresponding author)}
\affiliation{Department of Physics, Jiangxi Normal University, Nanchang 330022, China}

\date{\today}

\begin{abstract}
Recently, Fernandes discovered an analytic solution for rotating black holes in semiclassical gravity induced by the trace anomaly. These solutions exhibit some distinctive characteristics, including a non-spherically symmetric event horizon and violations of the Kerr bound. As a crucial assumption to uphold causality in spacetime, we investigate the validity of the weak cosmic censorship conjecture (WCCC) within this class of solutions with type-A trace anomaly by introducing a test particle on the equatorial plane. Our study reveals three distinct mechanisms that can potentially destroy the event horizon, leading to a violation of the WCCC. Our findings indicate that, with the exception of extremal Kerr, static extremal, and static singular black holes, the WCCC may be violated under the first-order perturbation of the test particle. These results suggest the need for further exploration of modifications to the behavior of the test particle under quantum effects in order to address the violation of the WCCC in this system.
\end{abstract}
\maketitle

\section{Introduction}

The Hawking-Penrose singularity theorem states that gravitational collapse inevitably ends up in a spacetime singularity \cite{Penrose:1964wq,Hawking:1970zqf}. However, these singularities in gravitational theories can lead to unpredictable results, making it difficult for us to understand the rules of the universe. Penrose suggested a solution known as the weak cosmic censorship conjecture (WCCC) \cite{Penrose:1969pc}, which proposes that these singularities must be hidden by an event horizon, maintaining the predictability of gravitational theories. Resolving this issue is crucial for understanding classical gravitational theory and could offer significant insights into the nature of the universe.

The WCCC has been tested in various ways, such as through numerical simulations with collapsing matter fields and disturbed black holes, and in simulations of merging black holes in higher dimensions \cite{Christodoulou:1984mz,Ori:1987hg,Shapiro:1991zza,Lemos:1991uz,Choptuik:1992jv,Corelli:2021ikv,Eperon:2019viw,Crisford:2017zpi,Figueras:2017zwa,
Figueras:2015hkb,Lehner:2010pn,Hertog:2003zs,Andrade:2020dgc,Andrade:2019edf,Andrade:2018yqu,Sperhake:2009jz}. In 1974, Wald designed a gedanken experiment \cite{Wald:1974}, which demonstrated that an extremal Kerr-Newman black hole could resist destruction from a test particle under the first-order approximation from the particle perturbation. Then, Hubeny expanded it to consider near-extremal black holes and second-order perturbations, suggesting these black holes could potentially be destroyed \cite{Hubeny:1998ga}. Many follow-up studies agreed with this finding \cite{deFelice:2001wj,Hod:2002pm,Jacobson:2010iu,Chirco:2010rq,Saa:2011wq,Gao:2012ca}. However, as Hubeny \cite{Hubeny:1998ga} discussed, to confirm whether black holes actually disintegrate, all second-order effects must be taken into account. In 2017, a more complex version of the gedanken experiment was proposed by Sorce and Wald based on the Noether charge method, which considered the full dynamics of spacetime and perturbation matters \cite{Sorce:2017dst} and showed that a near-extremal Kerr-Newman black hole cannot be destroyed under second-order perturbation when the matters satisfy the null energy condition. Additionally, field scattering is another method used to examine the WCCC across different gravitational systems \cite{Semiz:2005gs,Gwak:2018akg,Gwak:2021tcl,Liang:2020hjz,Natario:2016bay,Goncalves:2020ccm,Gwak:2019rcz,Yang:2020iat,Yang:2020czk,Feng:2020tyc,Yang:2022yvq}

The exploration of quantum phenomena offers us a deeper understanding of the universal laws of physics. Trace anomaly is one such phenomenon that emerges in a fundamentally conformally invariant classical theory due to the breaking of conformal symmetry by one-loop quantum corrections \cite{Capper:1974ic,Duff:1993wm}. This results in the renormalization of the stress-energy tensor, leading to a non-zero trace. Interestingly, this trace is independent of the quantum state of the quantum fields and solely depends on the local curvature of spacetime, marking it as a general characteristic of quantum theories in gravitational fields \cite{Anderson:2007eu,Mottola:2022tcn,Mottola:2006ew,Mottola:2016mpl}.

In a four-dimensional spacetime, the trace anomaly can be expressed in terms of the square of the Weyl tensor $C$ and the Gauss-Bonnet scalar $\mathcal{G}$, which are commonly referred to as type-A and type-B anomalies respectively \cite{Deser:1993yx}. This can be captured by the following equation,
\ba
g^{\mu \nu}\left\langle T_{\mu \nu}\right\rangle=\frac{\beta}{2} C^{2}-\frac{\alpha}{2} \mathcal{G}\,.
\ea
When contemplating modifications to General Relativity, the contributions of the trace anomaly are critical. They are anticipated to produce observable macroscopic effects \cite{Mottola:2022tcn,Mottola:2006ew,Mottola:2016mpl}, thereby necessitating their inclusion in the low-energy effective field theory of gravity.
\begin{figure*}
    \centering
    \includegraphics[width=0.7\textwidth]{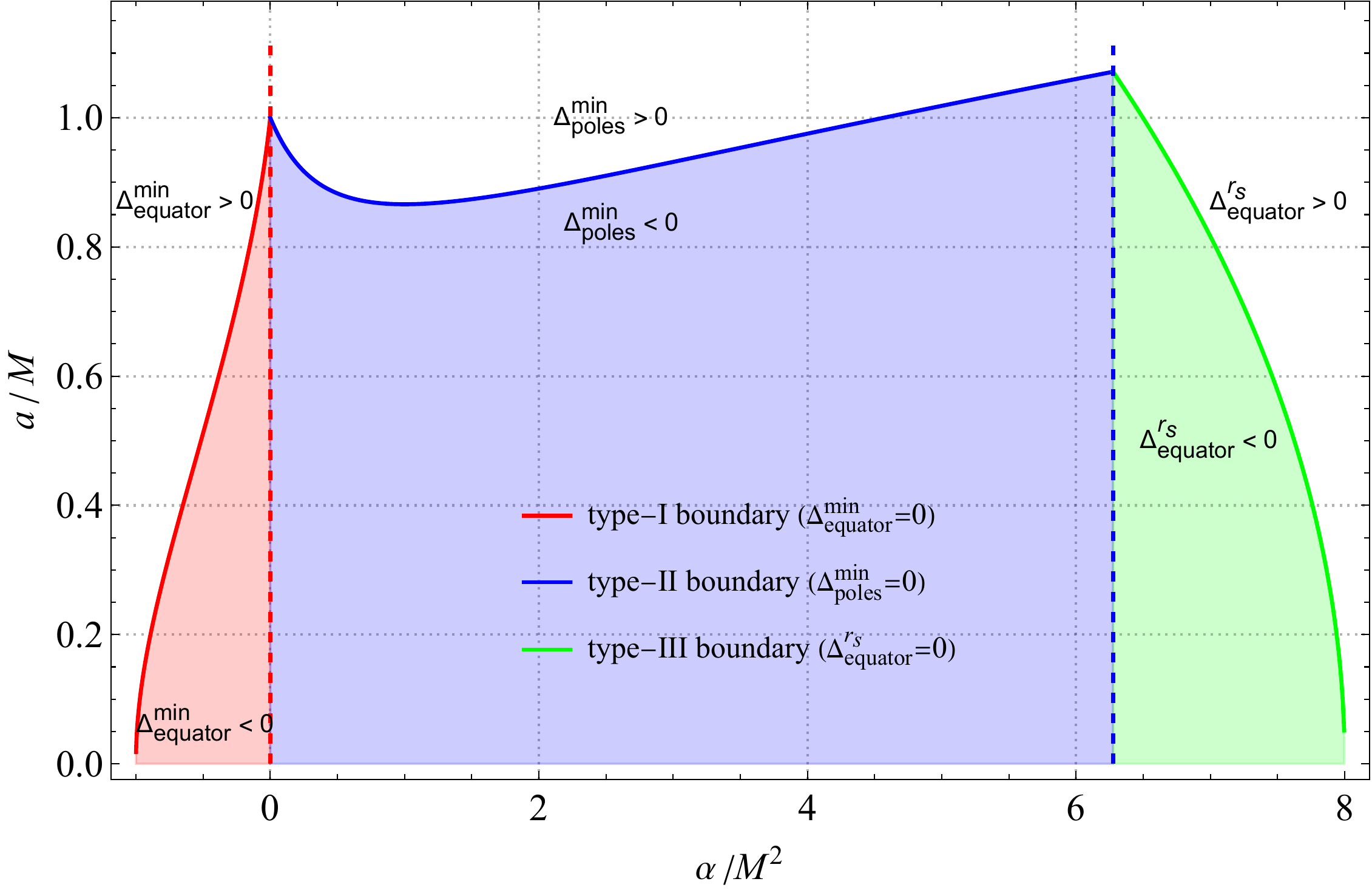}
    \caption{
The $\left(\alpha / M^{2}, a / M\right)$ diagram illustrating the domain of existence for black hole solutions. The shaded region represents the parameter space where black hole solutions exist. The boundary of this region is determined by three distinct boundaries. The type-I boundary (red) is defined by $\Delta_\text{equator}^\text{min}=0$, the type-II boundary (blue) is defined by $\Delta_\text{poles}^\text{min}=0$, and the type-III boundary (red) corresponds to the singular black hole solutions defined by $\Delta_\text{equator}^{r_s}=0$.}
    \label{fig1}
\end{figure*}

Utilizing the semi-classical approach allows us to account for the backreaction of quantum fields and their influence on spacetime geometry. This process transforms the Einstein equation into
\ba
R_{\mu \nu}-\frac{1}{2} g_{\mu \nu} R=8 \pi G\left\langle T_{\mu \nu}\right\rangle\,.
\ea
A notable challenge when studying the backreaction of quantum fields lies in the typically unknown expectation value of the renormalized stress-energy tensor even in the static and spherically symmetric system \cite{Christensen:1977jc,Ho:2018fwq,Abedi:2015yga}. To overcome this challenge, Ref. \cite{Cai:2009ua} introduces an additional condition: the geometry should rely solely on a single free function, which effectively establishes an additional equation of state for the stress-energy tensor. Through the adoption of this methodology, and focusing solely on the type-A anomaly $(\beta=0)$, Ref. \cite{Cai:2009ua} manages to fully derive the renormalized stress-energy tensor and obtain an analytic static and spherically symmetric black hole solution from the semi-classical Einstein equations. Intriguingly, these solutions show a logarithmic correction to their entropy, aligning with the expectation that primary quantum corrections to black hole entropy should be logarithmic \cite{Page:2004xp,Sen:2012dw}. Most recently, an analytic stationary and axially-symmetric black hole solution to the semiclassical Einstein equations induced by the trace anomaly has been found in Ref. \cite{Fernandes:2023vux}. Unlike conventional stationary black hole solutions, this new solution presents several distinct features, including the violation of the Kerr bound and an event horizon that lacks spherical symmetry, leading to a mismatch between the event horizon and the Killing horizon.

As a fundamental assumption for ensuring causality in spacetime, a natural question arises as to whether the WCCC still holds when considering quantum effects such as the trace anomaly. Therefore, the aim of this paper is to examine the WCCC in the rotating stationary solutions obtained in Ref. \cite{Fernandes:2023vux}. Specifically, we will investigate the possibility of destroying the event horizon by dropping a test particle into the black hole, thereby forming a naked singularity. The structure of this paper is as follows. In Sec. \ref{sec2}, we will introduce rotating black hole solutions in semiclassical gravity with type-A trace anomaly and discuss their spacetime structures. In Sec. \ref{sec3}, we will first present the equations of motion for test particles on the equatorial plane and derive the condition for test particle to enter the black hole. We will then discuss whether the black hole can be destroyed under the first-order approximation of the perturbation caused by the test particle, thus violating the WCCC. Finally, in Sec. \ref{sec4}, we will provide our conclusions and summarize the findings of our paper.

\section{Rotating black hole solutions in semiclassical gravity\label{sec2}}

In this paper, we focus exclusively on type-A anomalies (i.e., $\beta = 0$) in Einstein gravity. Most recently, Ref. \cite{Fernandes:2023vux} provides the solution to the semiclassical Einstein equations with trace anomaly, which corresponds to an asymptotically flat and rotating spacetime. In the ingoing Kerr-like coordinates $x^{\mu}=(v, r, \theta, \varphi)$, the metric is given by
\begin{equation}\label{dsmet}\begin{aligned}
d s^{2}= & -\left(1-\frac{2 \mathcal{M}(r, \theta) r}{\Sigma}\right)\left(d v-a \sin ^{2} \theta d \varphi\right)^{2} \\
& +2\left(d v-a \sin ^{2} \theta d \varphi\right)\left(d r-a \sin ^{2} \theta d \varphi\right) \\
& +\Sigma\left(d \theta^{2}+\sin ^{2} \theta d \varphi^{2}\right),
\end{aligned}\end{equation}
where
\begin{equation}\label{expressM}\begin{aligned}
\mathcal{M}(r, \theta)&=\frac{2M}{1 + \sqrt{1-{8 \alpha r \xi M}/{\Sigma^{3}}}}\,,\\
\Sigma&=r^{2}+a^{2} \cos ^{2} \theta\,,\\
\xi&=r^{2}-3 a^{2} \cos ^{2} \theta\,.
\end{aligned}\end{equation}
Here, after assuming the spacetime is asymptotically flat, the symbol $M$ represents an integration constant. The mass and angular momentum of the spacetime can be obtained by using the Komar integral, which yields $M$ and $J = M a$, respectively.

\begin{figure*}
    \centering
    \includegraphics[width=0.35\textwidth]{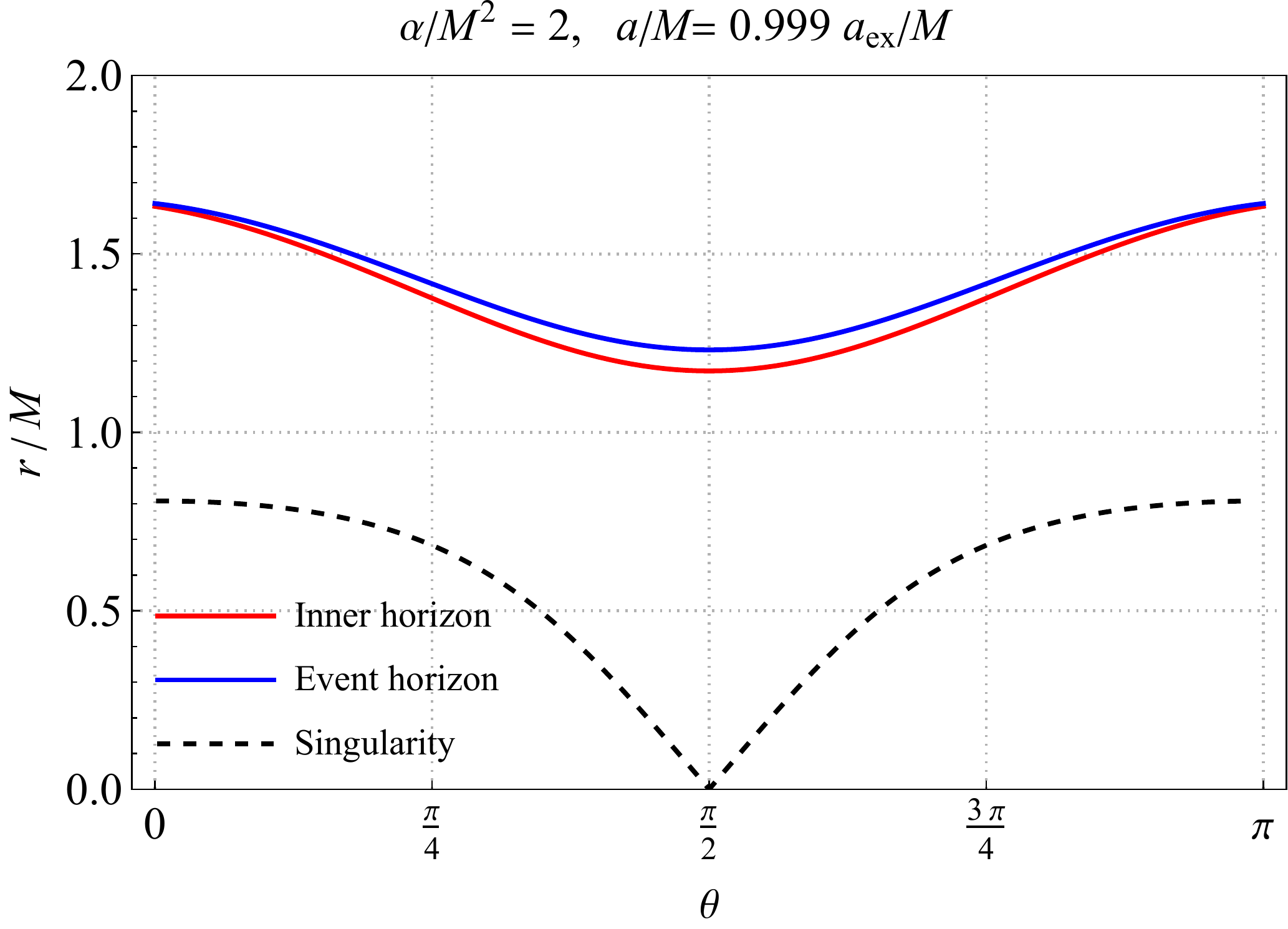}
    \includegraphics[width=0.35\textwidth]{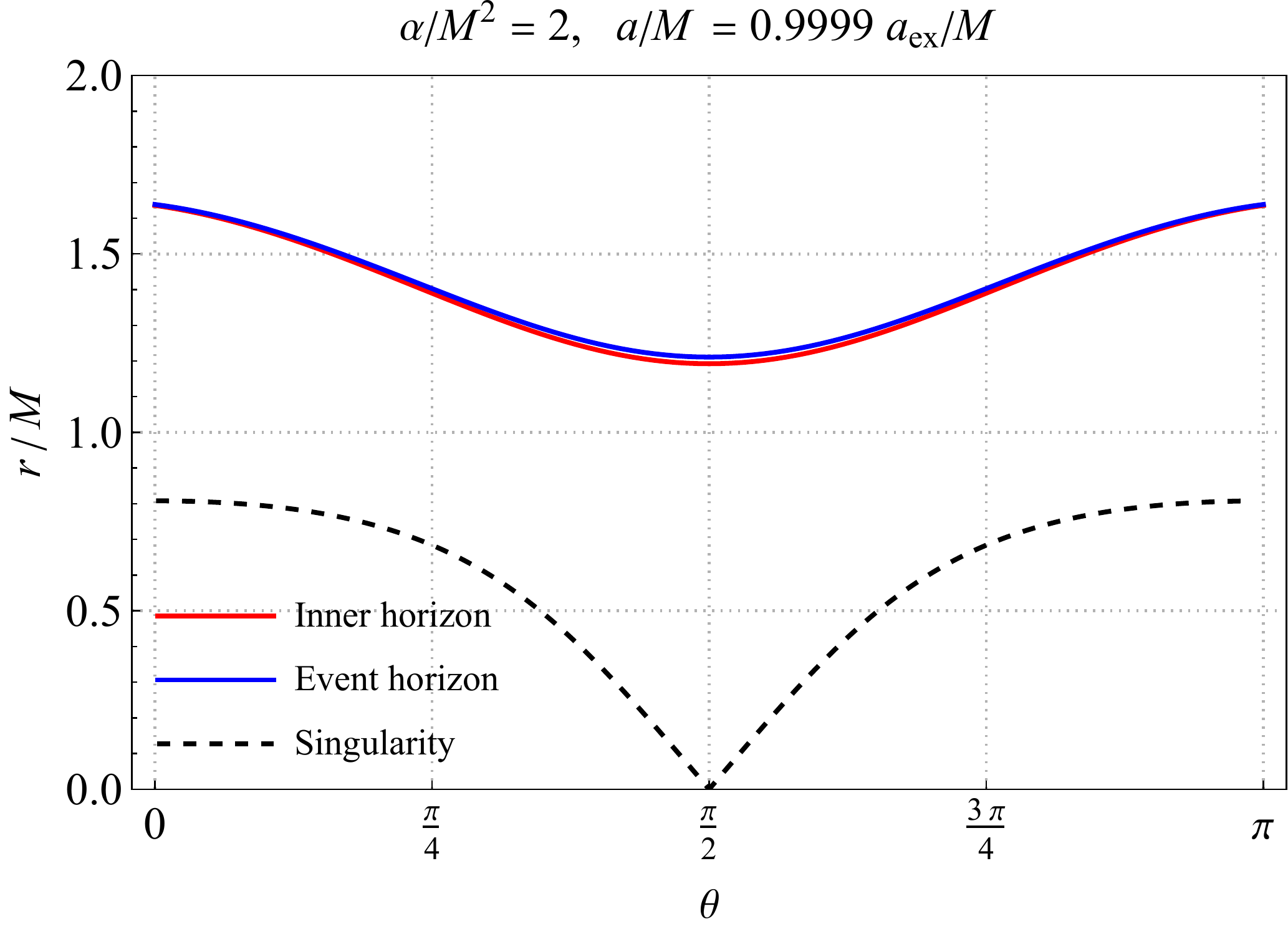}
    \includegraphics[width=0.35\textwidth]{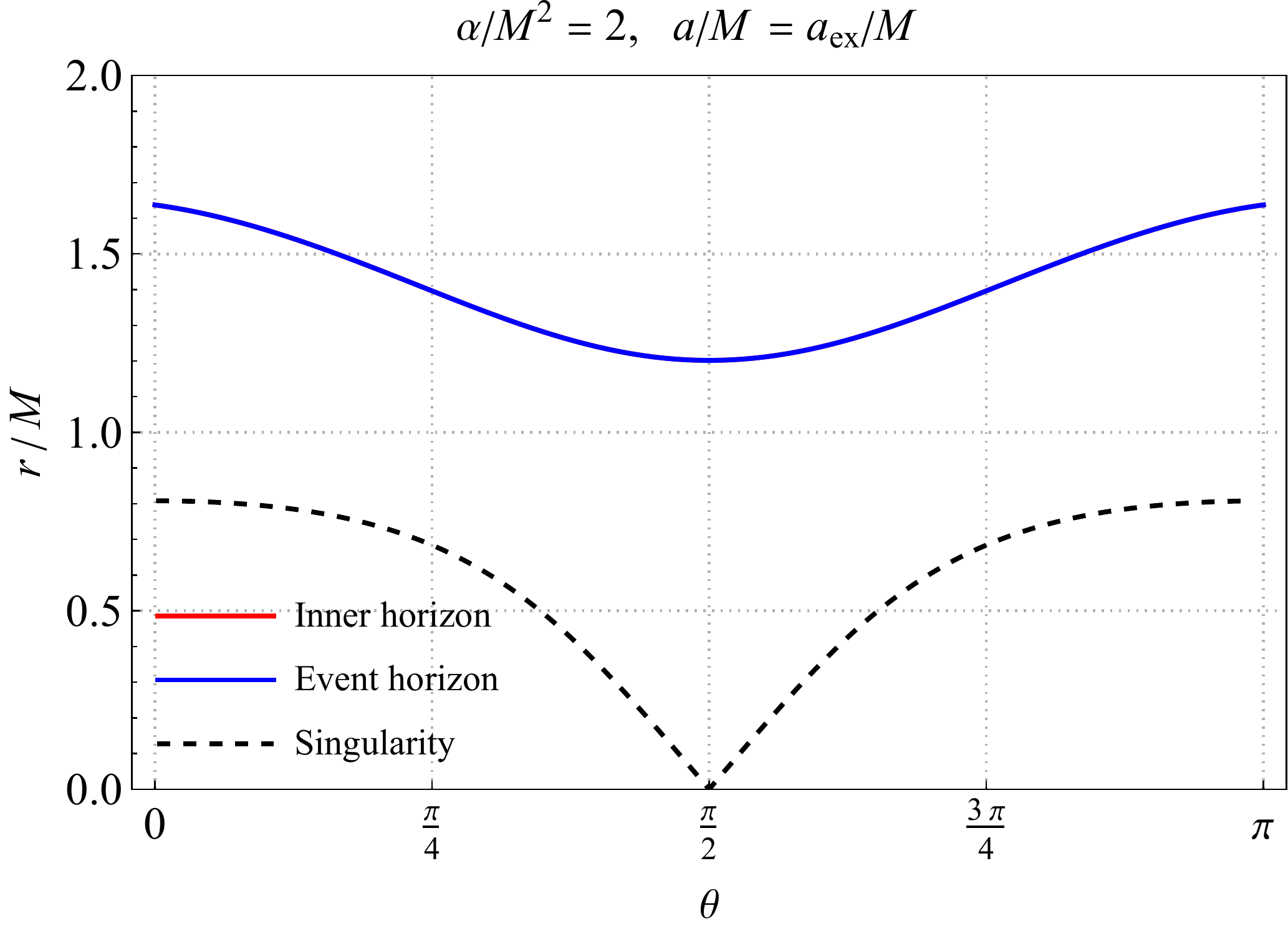}
    \includegraphics[width=0.35\textwidth]{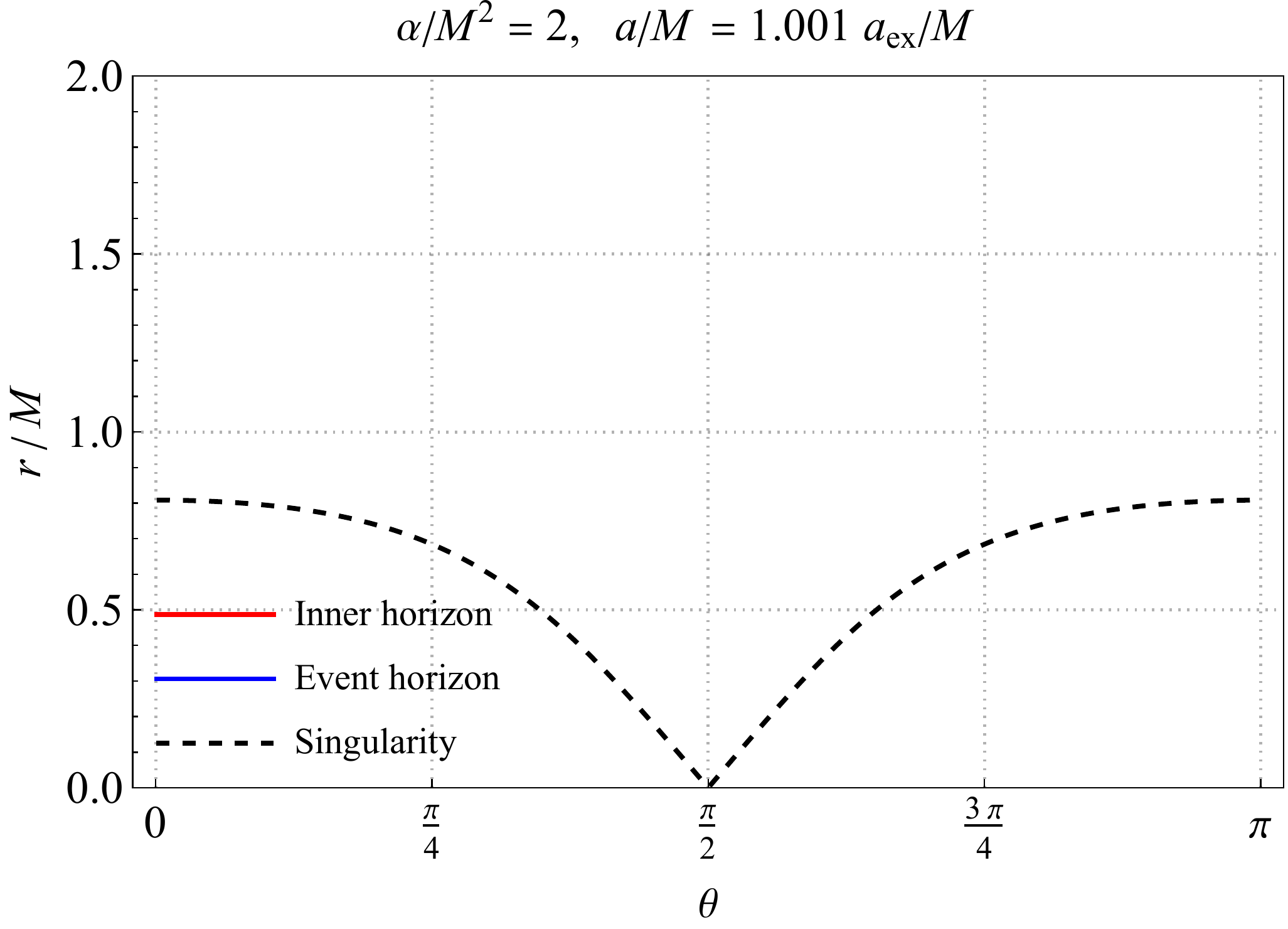}
    \caption{Diagrams illustrating the process of destroying an event horizon by increasing the angular momentum in parameter-region I ($-1\leq\a/M^2\leq0$). The profile shows the coordinate location of the inner horizon (represented by the red line) and the event horizon (represented by the blue line) as a function of the angular coordinate $\theta$. We consider a black hole with $\alpha / M^{2}=-0.3$ and varying values of $a/M^2$. Here, $a_\text{ex}/M = 0.667538$ corresponds to the extremal value of $a/M$ for $\alpha / M^{2}=-0.3$. The black dashed line indicates the position of the curvature singularity.}
    \label{fig2}
\end{figure*}
\begin{figure*}
    \centering
    \includegraphics[width=0.35\textwidth]{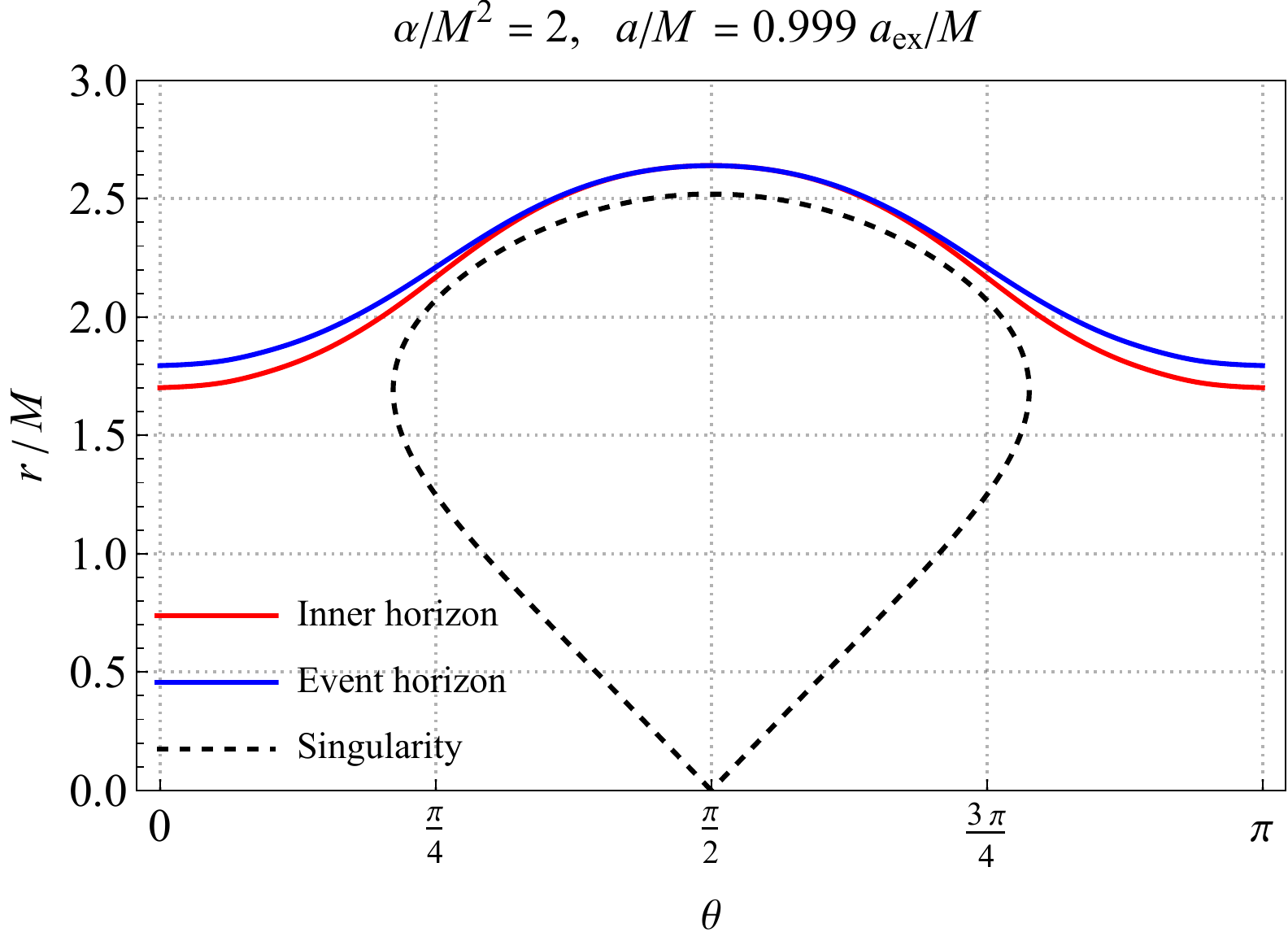}
    \includegraphics[width=0.35\textwidth]{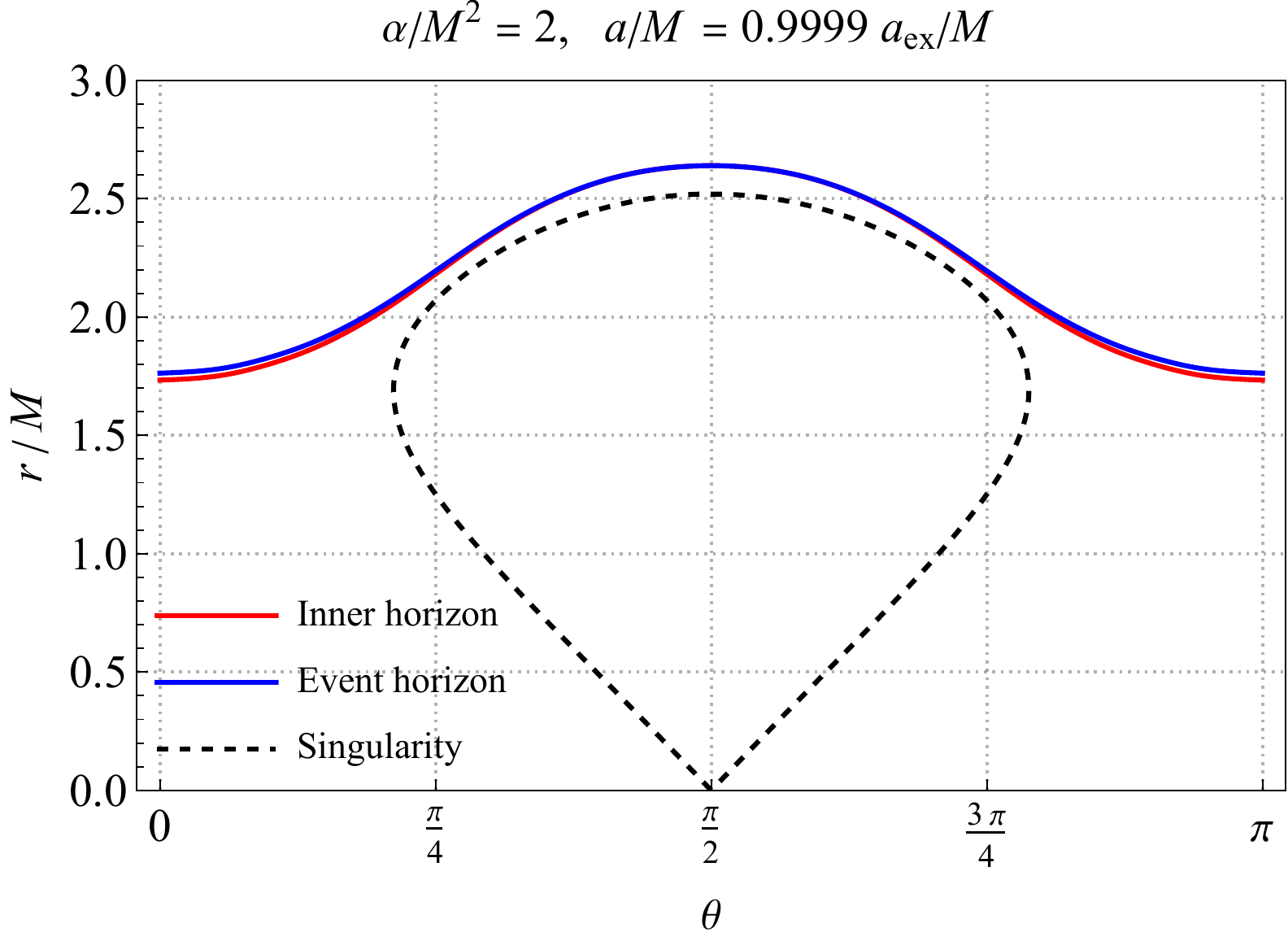}
    \includegraphics[width=0.35\textwidth]{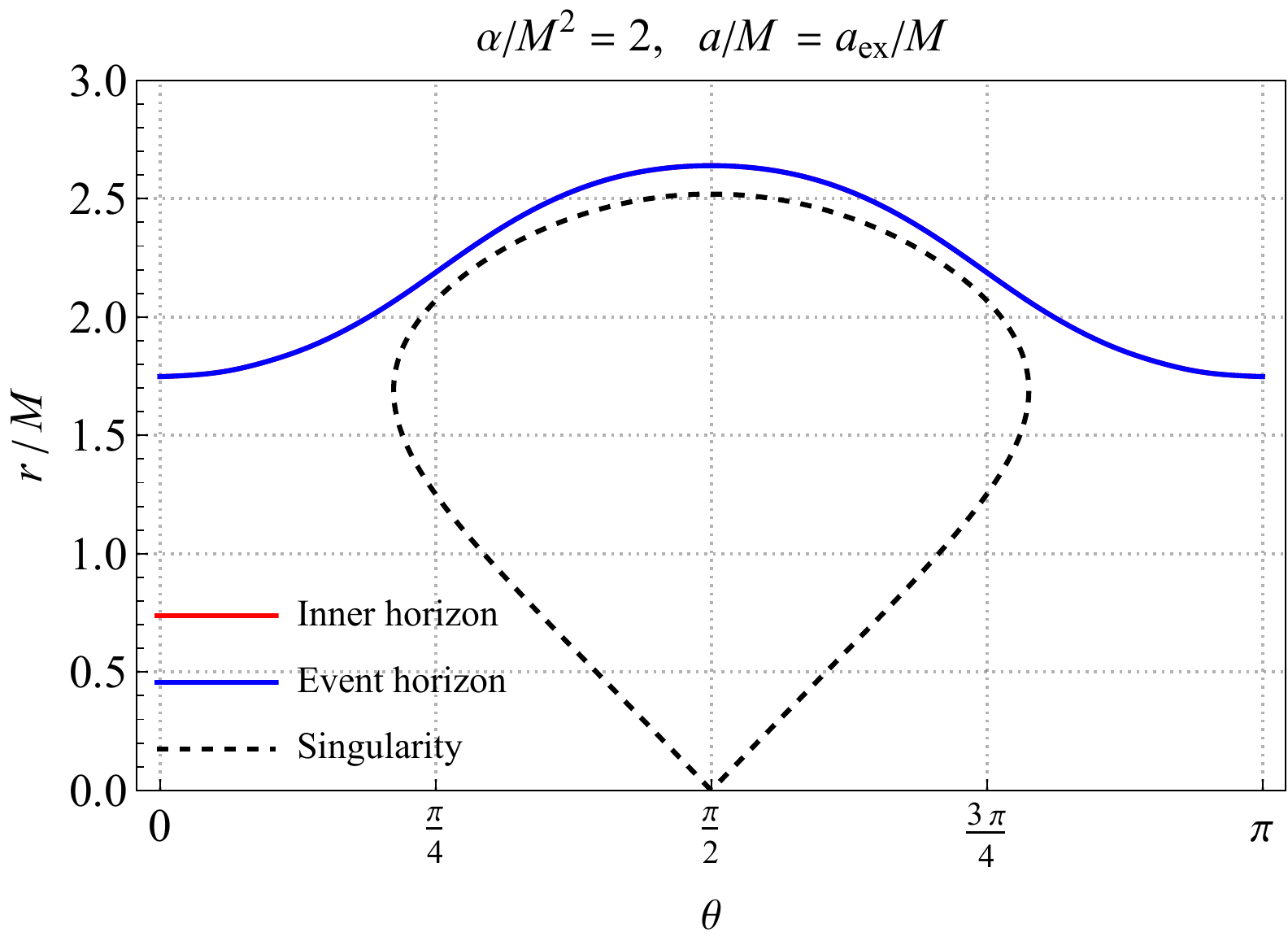}
    \includegraphics[width=0.35\textwidth]{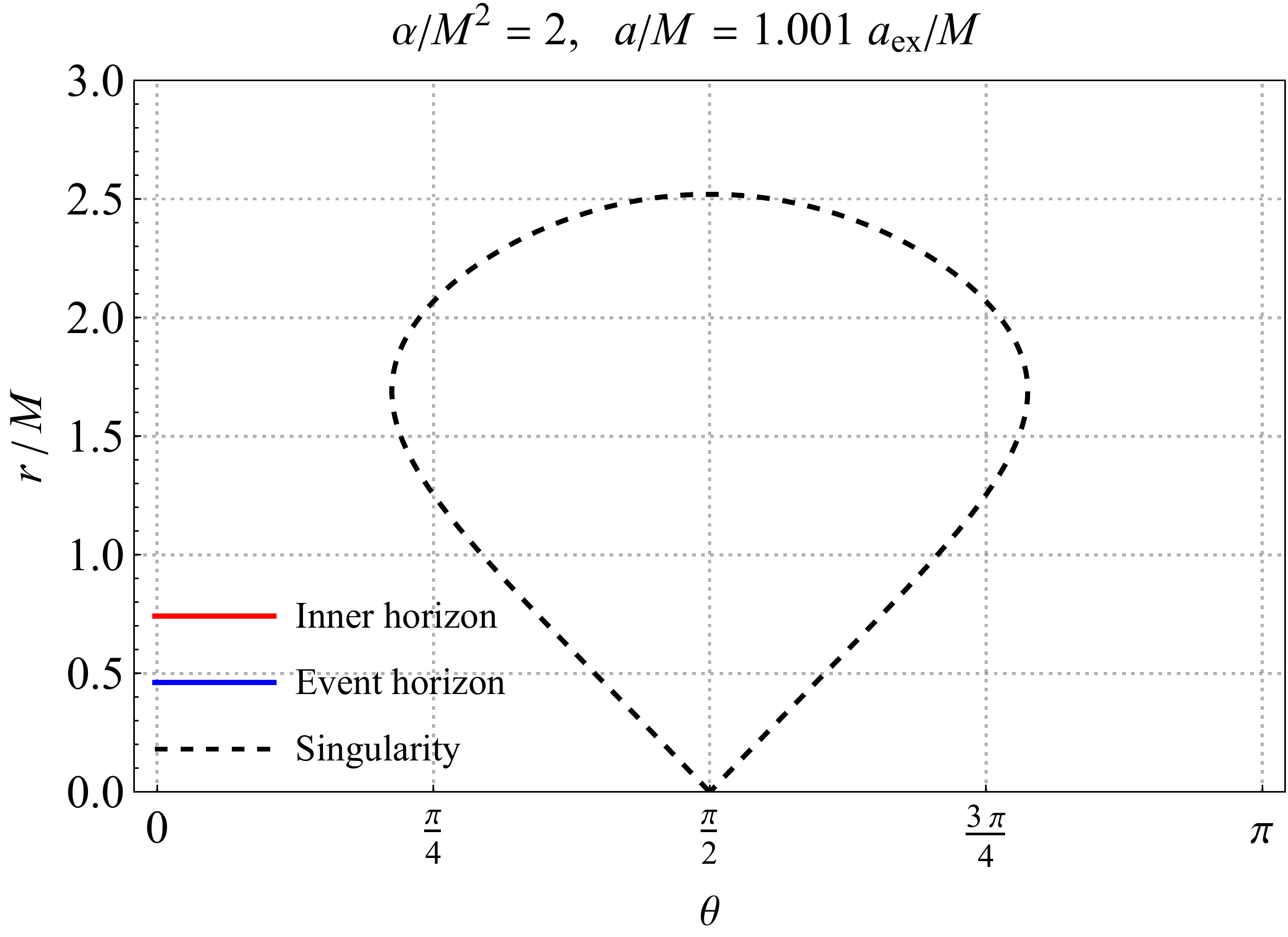}
    \caption{Diagrams illustrating the process of destroying an event horizon by increasing the angular momentum in parameter-region II ($(0 < \alpha/M^2 < 6.2754)$). The profile shows the coordinate location of the inner horizon (represented by the red line) and the event horizon (represented by the blue line) as a function of the angular coordinate $\theta$. We consider a black hole with $\alpha / M^{2}=2$ and varying values of $a/M^2$. Here, $a_\text{ex}/M = 0.890307$ corresponds to the extremal value of $a/M$ for $\alpha / M^{2}=2$. The black dashed line indicates the position of the curvature singularity.}
    \label{fig3}
\end{figure*}
\begin{figure*}
    \centering
    \includegraphics[width=0.35\textwidth]{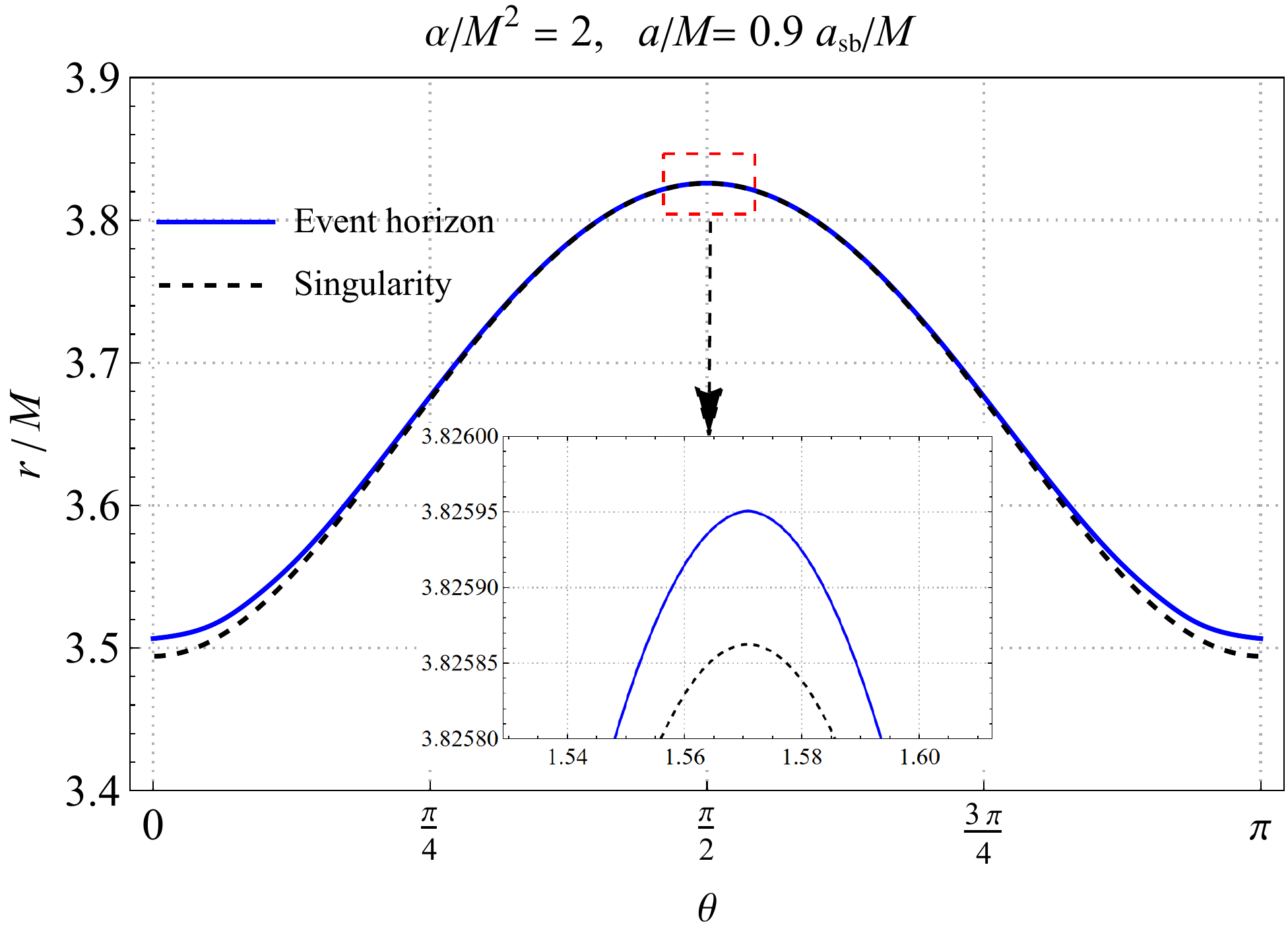}
    \includegraphics[width=0.35\textwidth]{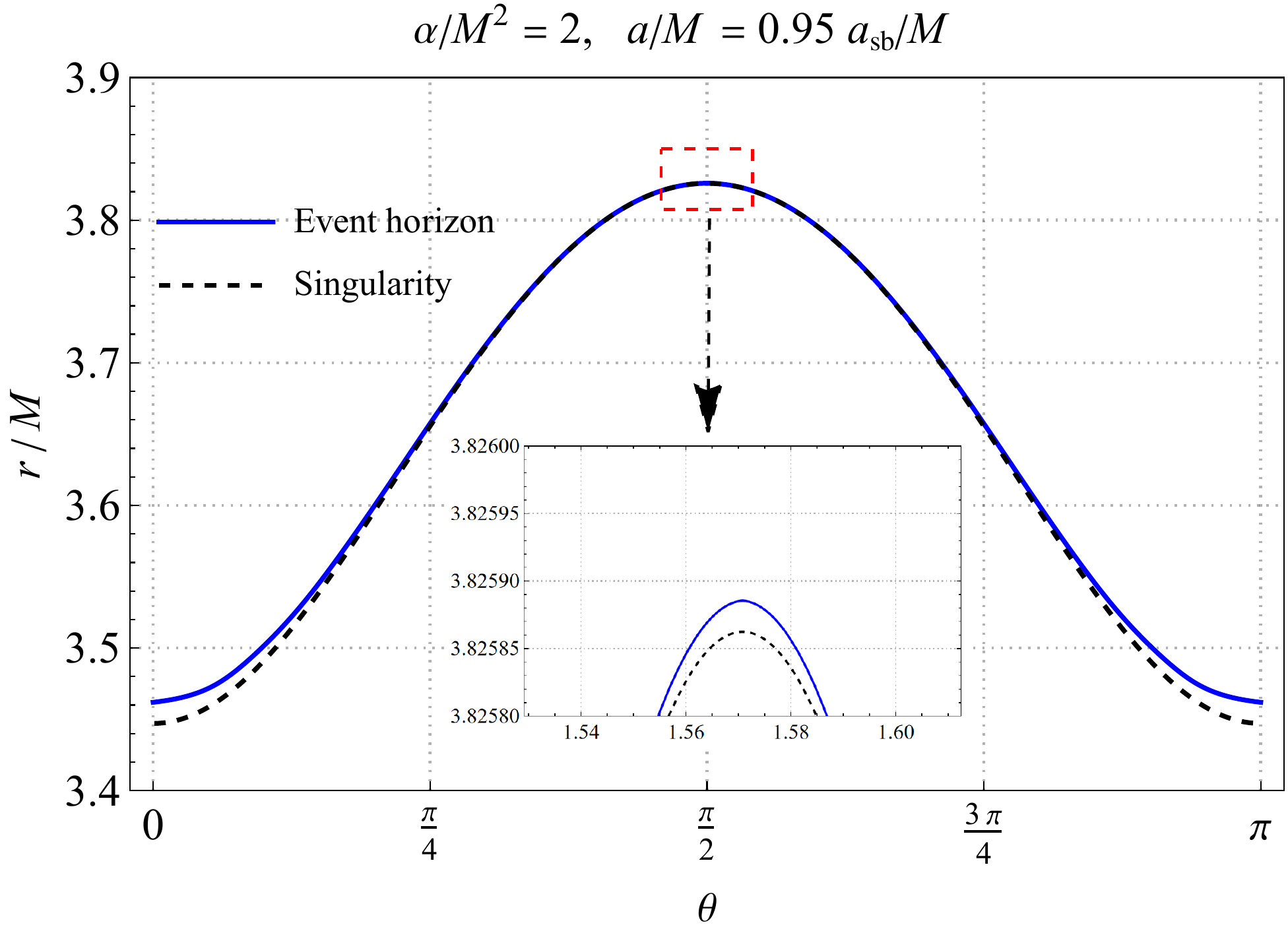}
    \includegraphics[width=0.35\textwidth]{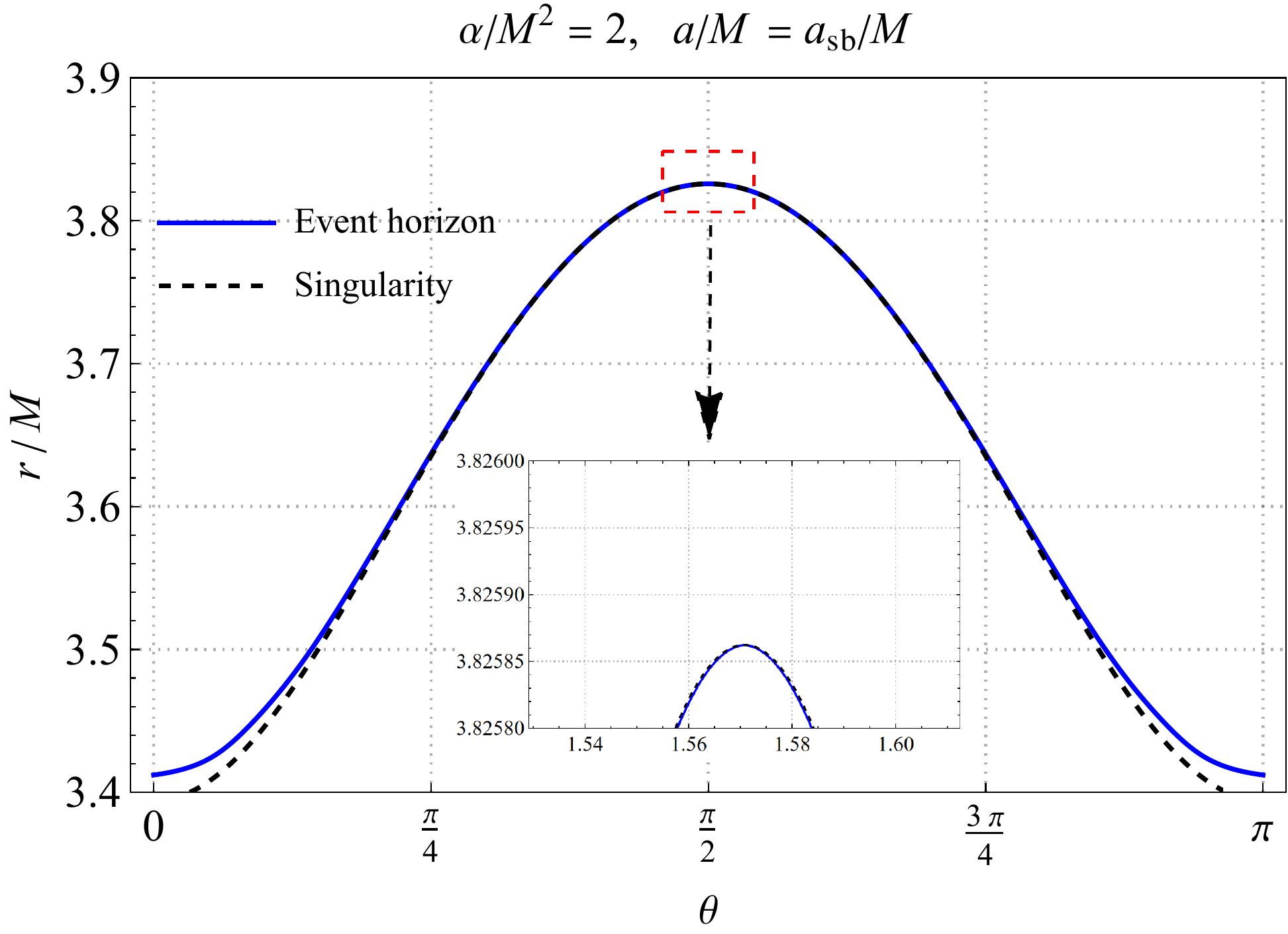}
    \includegraphics[width=0.35\textwidth]{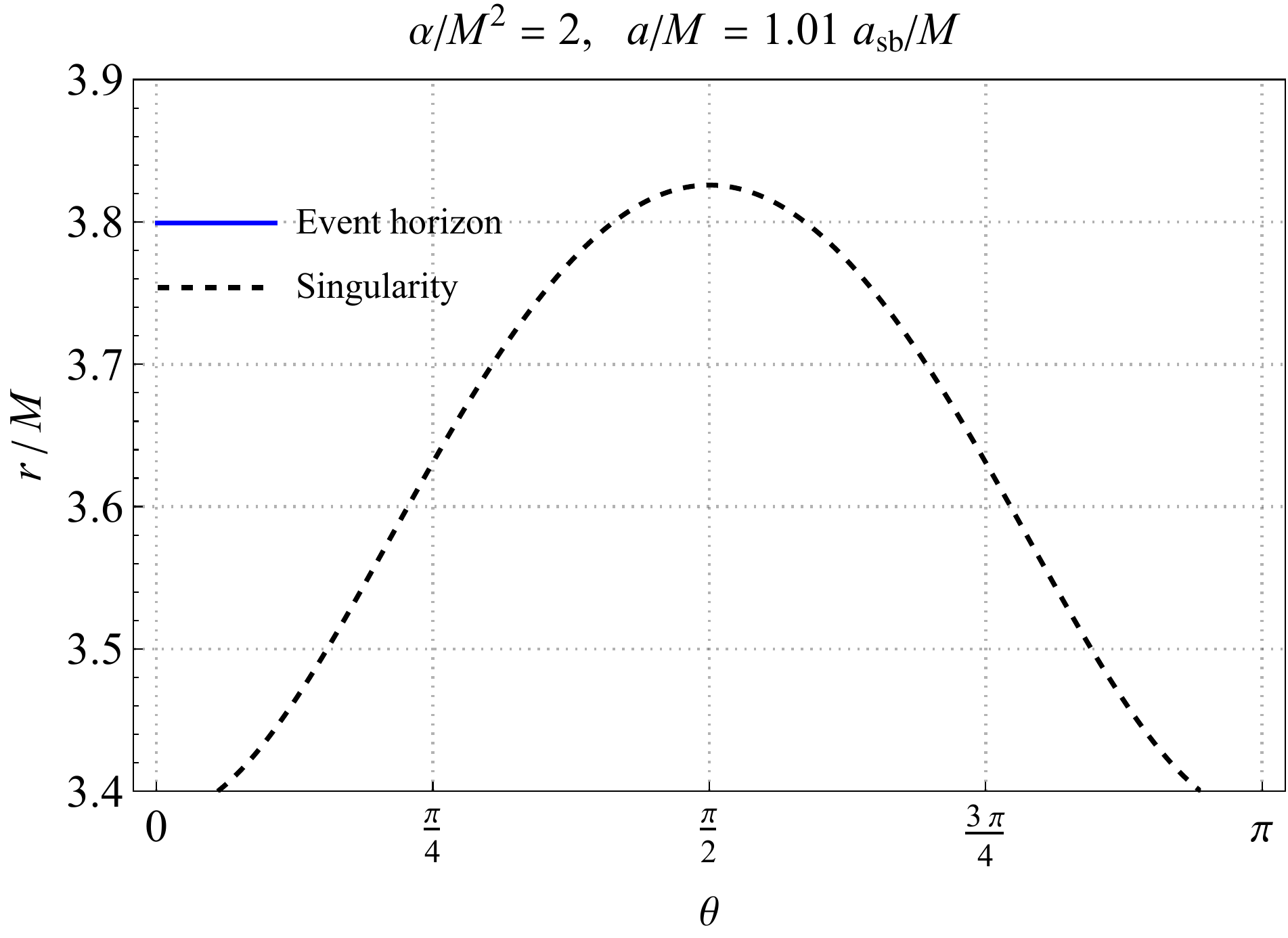}
    \caption{Diagrams illustrating the process of destroying an event horizon by increasing the angular momentum in parameter-region III ($6.2754\leq\a/M^2\leq 8$). The profile shows the coordinate location of the event horizon (represented by the blue line) and the singularity (represented by the black dashed line) as a function of the angular coordinate $\theta$. We consider a black hole with $\alpha / M^{2}=7$ and varying values of $a/M$. Here, $a_\text{sb}/M = 0.816227$ corresponds to the angular momentum of the singular black hole when $\alpha / M^{2}=7$. }
\label{fig4}
\end{figure*}

By analyzing the spacetime metric (\ref{dsmet}), we can identify the existence of two singularities. The first singularity is the well-known ring singularity, positioned at $\Sigma=0$, which corresponds to $r=0$ and $\theta=\pi / 2$. The second singularity arises due to the quantum effects of Gauss-Bonnet theory. It is located at the point $r=r_s(\theta)$ where the expression inside the square root in the mass function (\ref{expressM}) becomes zero, i.e.,
\ba
1-\frac{8 \alpha r \xi}{\Sigma^{3}} M=0\,.
\ea
Solving this condition requires a numerical approach, except for $\theta=\pi / 2$, where we find
\ba
r_{s}(\pi / 2)=2(M \alpha)^{1 / 3}\,.
\ea

In the context of describing a black hole spacetime, there exists an event horizon located at $r=r_H(\theta)$ that conceals the singularity within it. This event horizon satisfies the following differential equation
\ba\begin{aligned}\label{equationrh}
\left[\partial_{\theta} r_{H}(\theta)\right]^{2}+\left.\Delta\right|_{r=r_{H}(\theta)}=0\,,
\end{aligned}\ea
where
\begin{equation}
\Delta=r^2+a^2-2 r \mathcal{M}(r, \theta).
\end{equation}
Considering the symmetries of the problem, we need to establish the boundary conditions $\partial_{\theta} r_{H}(0)=\partial_{\theta} r_{H}(\pi / 2)=0$. Together with Eq. (\ref{equationrh}), these boundary conditions imply
\ba\begin{aligned}
\Delta_\text{poles}|_{r=r_H(0)} = 0\,,\quad\text{and}\quad \Delta_\text{equator}|_{r=r_H(\pi/2)}=0\,,
\end{aligned}\ea
in which we define
\ba\begin{aligned}
\Delta_\text{poles}(r)&\equiv \Delta(r, \theta =0, \pi)\,,\\
\Delta_\text{equator}(r)&\equiv \Delta(r, \theta =\pi/2)\,.
\end{aligned}\ea
To establish the presence of an event horizon, it is necessary for both $\Delta_\text{poles}(r)$ and $\Delta_\text{equator}(r)$ to have at least one root. Solving Eq. \eq{equationrh} using the pseudospectral method enables us to determine the range of black hole existence, illustrated in Fig. \ref{fig1}. The figure reveals three distinct boundary behaviors within this range.

First, in the parameter-region I $(-1 \leq \alpha/M^2 \leq 0)$, the boundary of the black hole solution is represented by the red curve in Fig. \ref{fig1}. They describe extremal black hole solutions where the event horizon and the inner horizon coincide at the equator  ($\theta=\pi/2$), see Fig. \ref{fig2}. In this case, $\Delta_\text{equator}(r_H(\p/2))=\partial_r\Delta_\text{equator}(r_H(\p/2))=0$, or the minimum value of $\Delta_\text{equator}(r)$, denoted as $\Delta_\text{equator}^\text{min}$, is zero. Increasing the angular momentum further results in the absence of positive roots for $\Delta_\text{equator}(r)$ (i.e., $\Delta_\text{equator}^\text{min} > 0$), leading to the destruction of the event horizon and the exposure of the singularity.

Second, in the parameter-region II $(0 < \alpha/M^2 < 6.2754)$, the boundary of the black hole solutions is represented by the blue curve in Fig. \ref{fig1}. They describe extremal black hole solutions where the event horizon and the inner horizon overlap at  $\theta =0$ and $\pi$, see Fig. \ref{fig3}. In this case, the minimum value of $\Delta_\text{poles}(r)$, denoted as $\Delta_\text{poles}^\text{min}$, is zero. Further increasing the angular momentum leads to $\Delta_\text{poles}^\text{min} <0 $, resulting in the destruction of the event horizon.

Finally, in the parameter-region III $(6.2754 \leq \alpha/M^2 \leq 8 )$, we find the boundary of the third type showed by the green curve in Fig. \ref{fig1}, where the event horizon and the singularity overlap, see Fig. \ref{fig4}. Specifically, this boundary is determined by the equation $r_H(\pi/2) = r_s(\pi/2)=2(M\alpha)^{1/3}$, which also implies
\ba\begin{aligned}
\Delta_\text{equator}^{r_s}\equiv \Delta_\text{equator}|_{r=r_s(\pi/2)}=0\,.
\end{aligned}\ea

\section{Destroy the event horizon by throwing a test particle}\label{sec3}

In this section, our goal is to investigate whether the event horizon of a rotating black hole in semiclassical gravity with type-A anomaly can be destroyed after absorbing a test charged particle. The equation of motion for the test particle is given by
\ba\begin{aligned}
U^b\grad_b U^a=0,
\end{aligned}\ea
where
\ba\begin{aligned}
U^a=\left(\frac{\partial}{\partial \tau}\right)^a
\end{aligned}\ea
represents the four-velocity of the particle, and $\tau$ denotes the particle's proper time, ensuring $U^a U_a=-1$. The energy $E$ and angular momentum $L$ of the test particle are defined as
\ba\begin{aligned}\label{EL}
E&\equiv-m U_a \left(\frac{\partial}{\partial v}\right)^a=-m U_v\,,\\
L&\equiv m U_a \left(\frac{\partial}{\partial \varphi}\right)^a=m U_\varphi\,.
\end{aligned}\ea
For simplicity and without loss of generality in our subsequent analysis, we set $m=1$. To treat the particle as a test body, we assume that its energy $E$ and angular momentum $L$ are small compared to those of the black hole, i.e.,
\ba\label{testcd}
E\ll M\,,\quad\text{and}\quad L \ll J\,.
\ea

Considering the reflection symmetry of the black hole with respect to the equatorial plane, we focus on the scenario where the particle moves on the equatorial plane $(\theta=\pi / 2)$ with some angular momentum, resulting in zero components of velocity $U_{\theta}=0$. Using the normalized condition $U^a U_a=-1$ and the definitions of energy and angular momentum, we can obtain the following expressions,
\ba\begin{aligned}
\left(\frac{d r}{d\tau}\right)^2 &= \left.\frac{a^2\left( E-\Omega_H L\right)^2-[r_H^2+(a E-L)^2]\Delta}{r_H^4\Omega_H^2}\right|_{\theta=\pi/2}\,,\\
\frac{d v}{d\tau}&= \left.\frac{r_H^2(a^2+r_H^2)+\mathcal{O}(E, L)}{2r_H^2 (a^2 E- a L+E r_H^2)}\right|_{\theta=\pi/2}\,,
\end{aligned}\ea
in which
\ba\begin{aligned}
\Omega_H\equiv \frac{a}{a^2+r_H^2}
\end{aligned}\ea
is the angular velocity of the Killing horizon $r=r_K(\theta)$, determined by $\Delta(r, \theta)=0$.

The condition for the test particle to enter the event horizon on the equatorial plane requires its motion near the event horizon $r=r_H$ to be timelike and future-directed, which implies
\ba\label{lowbound}
E > \Omega_H^\text{equator} L
\ea
with
\ba\begin{aligned}
\Omega_H^\text{equator} \equiv \Omega_H|_{\theta=\pi/2}\,.
\end{aligned}\ea
Here we have used the assumption (\ref{testcd}). The above inequality gives the lower bound of $E/L > \Omega_H^\text{equator}$ such that the test particle can enter the event horizon on the equatorial plane.

Next, we explore the conditions required to destroy the black hole. Specifically, the test particle must be capable of entering the event horizon, and the black hole should become overspun after absorbing the test particle. These conditions establish a relationship between the energy $E$ and angular momentum $J$ of the test particle. After the test particle is dropped into the black hole, the parameters of the final state become
\ba\begin{aligned}
M &\rightarrow M^{\prime}= M+\delta M, \\
J &\rightarrow J^{\prime}=J+\delta J,
\end{aligned}\ea
where $\delta M = E$ and $\delta J = L$. To examine the validity of the WCCC, we need to determine whether the spacetime with mass $M'$ and $J'$ still represents a black hole solution, i.e., whether $(M', J')$ lies within the domain of existence of black holes as shown in Fig. \ref{fig1}.

In the previous section, we showed that there are three different types of parameter regions, each associated with distinct mechanisms for the destruction of the event horizon. In the following, we will discuss the conditions for the destruction of black holes in each parameter region and analyze the possibility of black hole destruction when considering the particle conditions mentioned earlier.

\begin{figure*}
    \centering
    \includegraphics[width=0.355\textwidth]{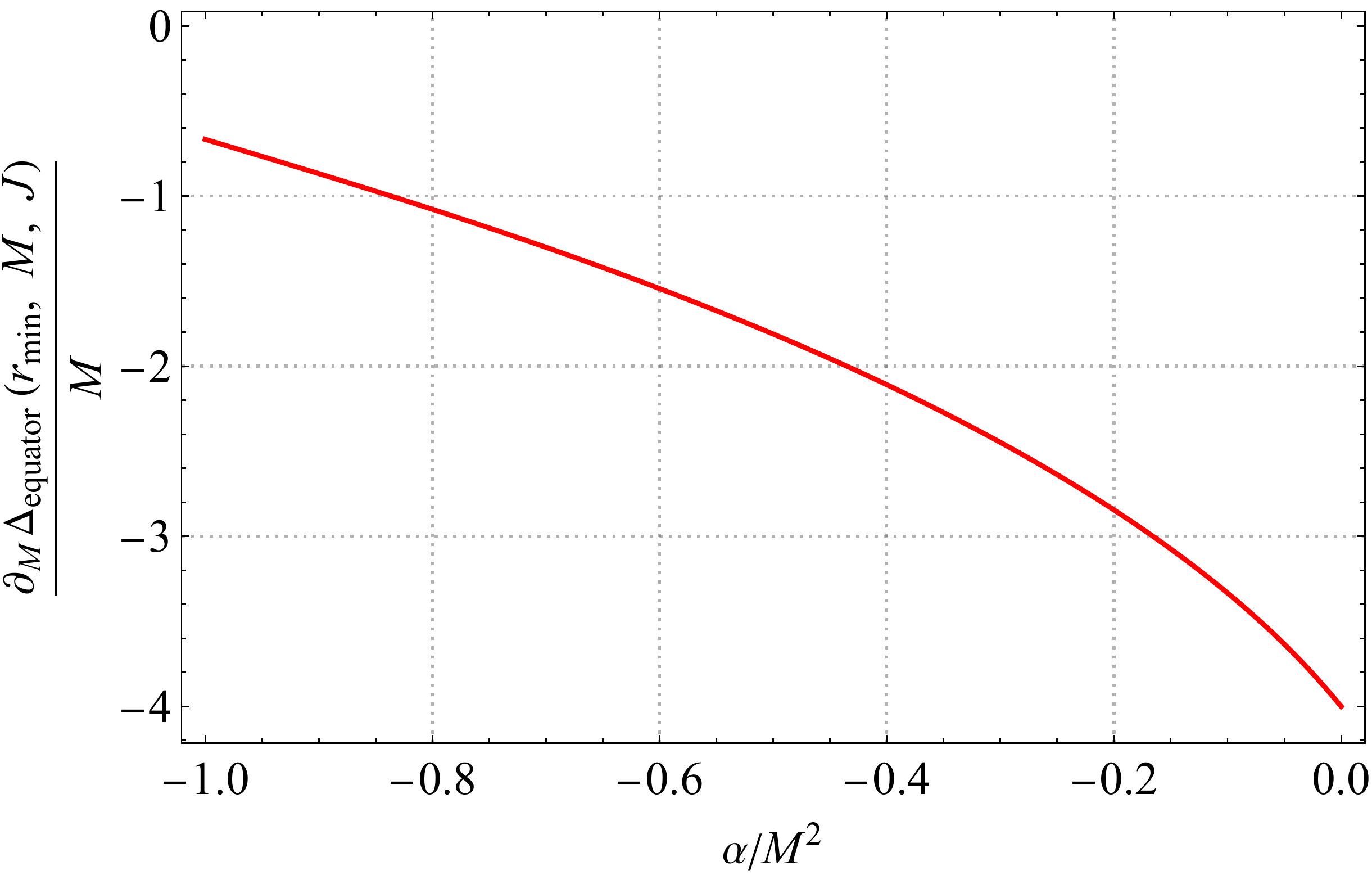}
    \includegraphics[width=0.35\textwidth]{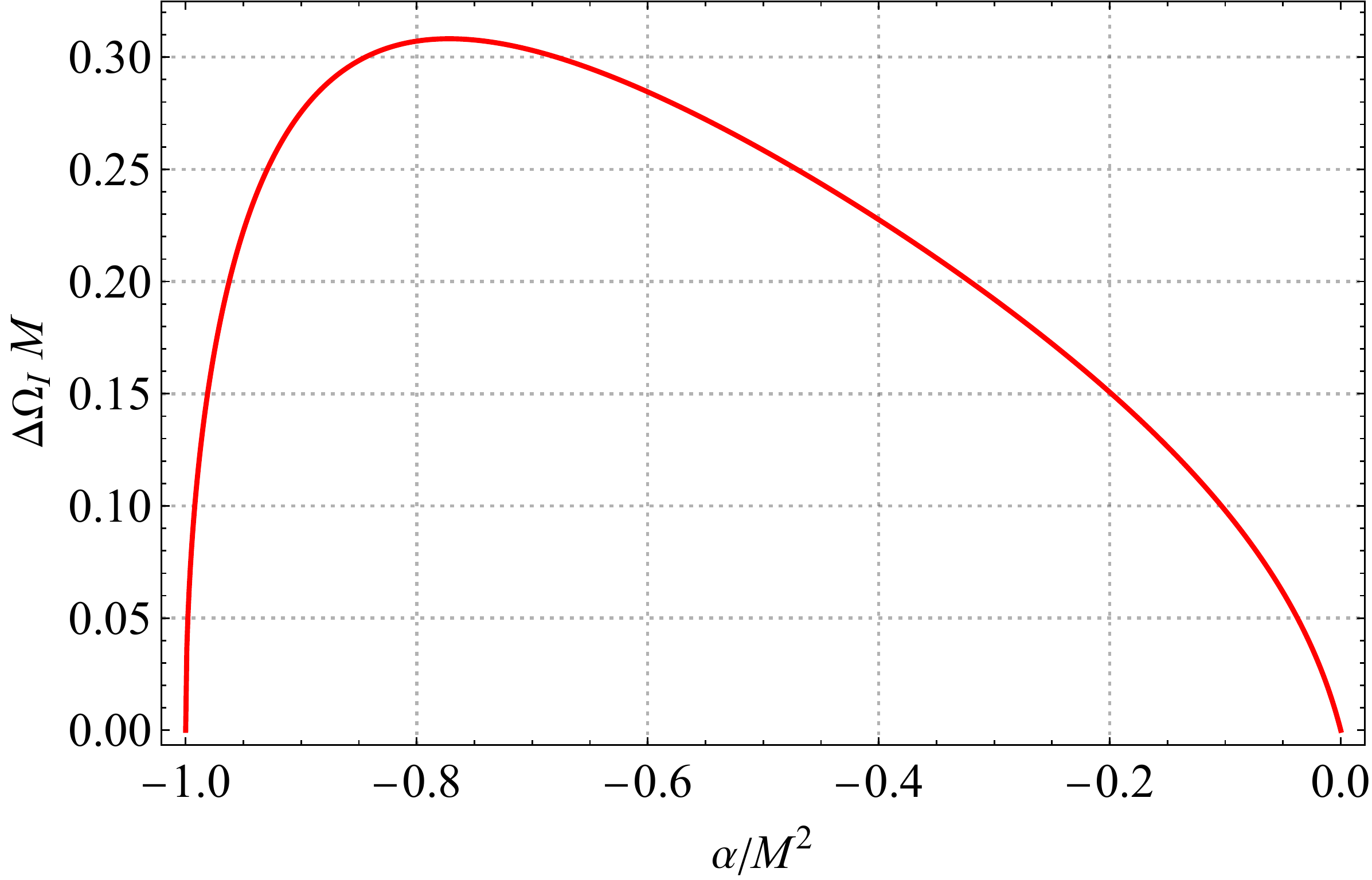}
    \caption{ $\partial_M \Delta_\text{equator}(r_\text{min}, M, J)$ and $\Delta \Omega_\text{I}$ of the extremal black hole solution as a function of the coupling constant in the parameter-region I. }
    \label{figI}
\end{figure*}

\subsection{Type-I violation}

Firstly, let's consider the case where the coupling constant satisfies $-1\leq\alpha/M^2\leq 0$, which corresponds to parameter-region I. In this region, the boundary curve of the black hole solutions represents extremal black holes where the event horizon and the inner horizon overlap at the equator $\theta=\pi/2$. To illustrate the mechanism for destroying the event horizon and exposing the singularity,  in Fig. \ref{fig2}, we depict the profiles of the inner horizon, outer horizon, and singularity in $(\theta, r/M)$ coordinates with $\alpha/M^2=-0.3$ and different angular momenta near the boundary line. From the figure, we observe that by increasing the angular momentum, the inner and event horizons will tend to coincide at the equator $\theta=\pi/2$, and ultimately, the event horizon is destroyed when the angular momentum exceeds its extremal value $a=a_\text{ex}$. We refer to this phenomenon as a type-I violation of the black hole in our study. Therefore, we can use $\Delta_\text{equator}^\text{min}$ to determine whether the spacetime describes a black hole in parameter-region I. Through numerical calculations, we find that the boundary (red line in Fig. \ref{fig1}) is determined by $\Delta_\text{equator}^\text{min} = 0$, $\Delta_\text{equator}^\text{min} < 0$ in the red shaded region in Fig. \ref{fig1}, and $\Delta_\text{equator}^\text{min} > 0$ in the white region of Fig. \ref{fig1}. Hence, the destruction of the black hole after dropping a test particle in this parameter region implies that
\begin{equation}\label{dscdI}
\Delta'{}_\text{equator}^\text{min} > 0,
\end{equation}
where $\Delta'{}_\text{equator}^\text{min}$ is the minimal value of
\begin{equation}
\Delta_\text{equator}(r, M+\delta M, J+\delta J)\equiv\Delta(r, \pi/2, M+\delta M, J+\delta J).
\end{equation}

Next, we consider only the first-order approximation from the test particle, and the initial state of the spacetime is an extremal black hole, i.e., the solution lies on the red boundary curve in Fig. \ref{fig1}. Let $r_\text{min}$ be the point of minimal value for the function $\Delta_\text{equator}(r, M, J)$, meaning that we have $\Delta_\text{equator}^\text{min}=\Delta_\text{equator}(r_\text{min}, M, J)$. The condition for the initial state to be extremal implies
\begin{equation}\label{extremalcd1}
\Delta_\text{equator}(r_\text{min}, M, J)=0,
\end{equation}
which also implies that $r_\text{min}=r_H(\pi/2)$.

After dropping a test particle, the minimal point shifts infinitesimally to $r_{\text{min}}+\delta r_{\text{min}}$. For infinitesimal changes, we have
\begin{equation}\label{equation1111}\begin{aligned}
\Delta'{}_\text{equator}^\text{min} &= \Delta_\text{equator}(r_{\text{min}}+\delta r_{\text{min}}, M+\delta M, J+\delta J)\\
 &= \frac{\partial \Delta_\text{equator}}{\partial M}\delta M+\frac{\partial \Delta_\text{equator}}{\partial J}\delta J,
\end{aligned}\end{equation}
under the first-order approximation of the particle perturbation. Here, we have used the assumption (\ref{extremalcd1}) that the initial state is extremal and the condition that $r_\text{min}$ is the point of minimal value of $\Delta_\text{equator}(r, M, J)$, i.e.,
\begin{equation}
\partial_r\Delta_\text{equator}(r_\text{min}, M, J)=0.
\end{equation}
In the left panel of Fig. \ref{figI}, we demonstrate that
\begin{equation}
\partial_M\Delta_\text{equator}(r_\text{min}, M, J)<0
\end{equation}
for the extremal black hole solutions in parameter-region I, where $-1\leq\alpha/M^2\leq 0$. Then, the destruction condition (\ref{dscdI}), together with Eq. (\ref{equation1111}), yields
\begin{equation}
\delta M < \Omega_\text{I} \delta J,
\end{equation}
where we define
\begin{equation}
\Omega_\text{I}=-\frac{\partial_J \Delta_\text{equator}(r_\text{min}, M, J)}{\partial_M \Delta_\text{equator}(r_\text{min}, M, J)}.
\end{equation}
This provides an upper bound for $E/L=\delta M/\delta J$. Together with the condition (\ref{lowbound}) that the particle can be dropped into the black hole on the equatorial plane, the allowed range of $E/L$ for destroying the black hole is
\begin{equation}
\Omega_H^\text{equator}<E/L<\Omega_\text{I}.
\end{equation}
In the right panel of Fig. \ref{figI}, we show the allowed length
\begin{equation}
\Delta\Omega_\text{I} \equiv \Omega_\text{I}-\Omega_H^\text{equator}
\end{equation}
of $E/L$ as a function of the coupling constant $\alpha/M^2$ in parameter-region I. Consequently, we observe that the allowed length $\Delta\Omega$ is positive in the parameter region $-1 < \alpha/M^2<0$, and $\Delta\Omega = 0$ for $\alpha/M^2=0$ and $\alpha/M^2=-1$. The above results indicate that, except for the extremal Kerr limit ($\alpha/M^2=0$) and static limit ($\alpha/M^2=-1$) cases, the extremal black hole can be overspun by throwing a test particle. In other words, the WCCC is violated in these cases.

\begin{figure*}
    \centering
    \includegraphics[width=0.371\textwidth]{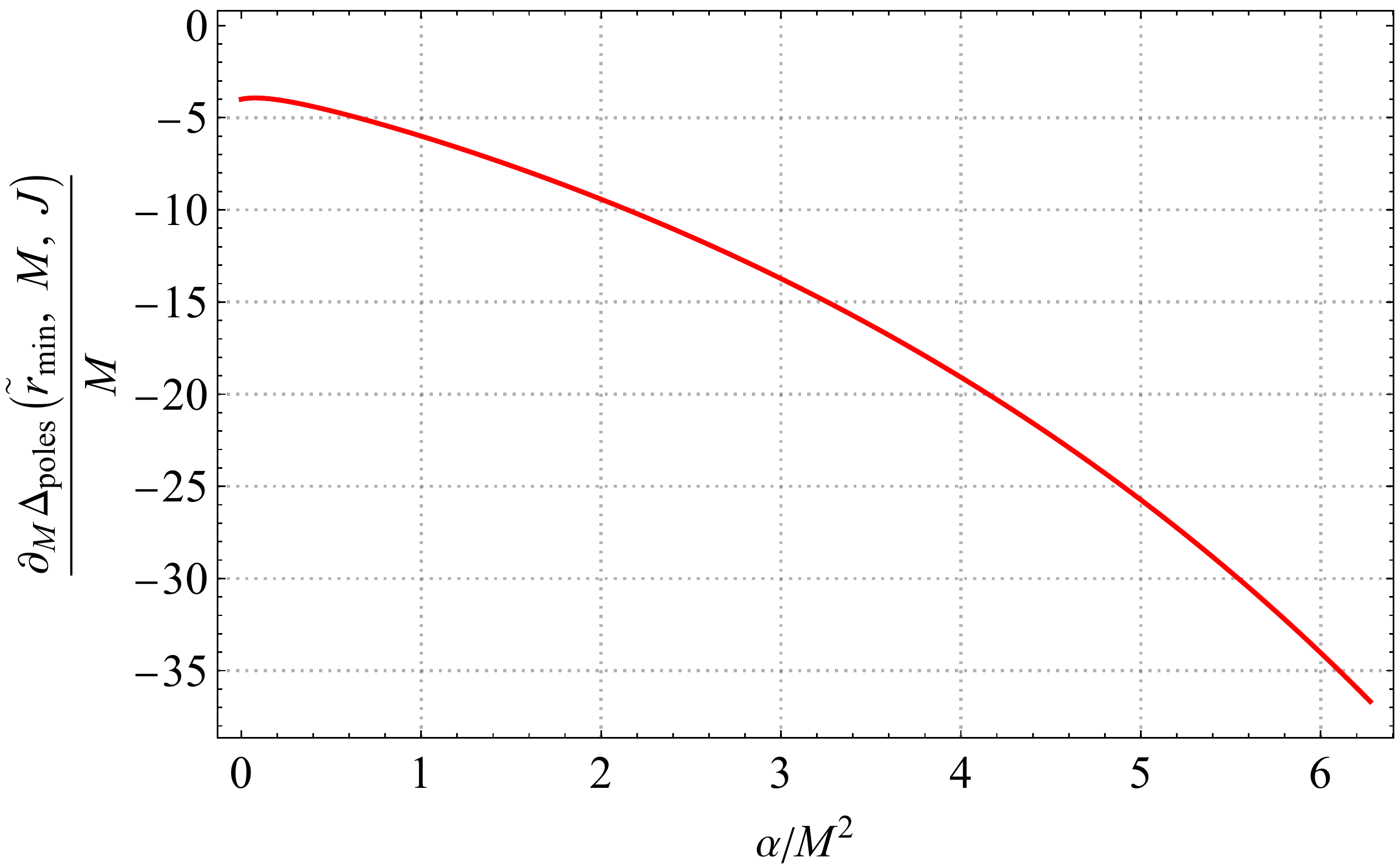}
    \includegraphics[width=0.35\textwidth]{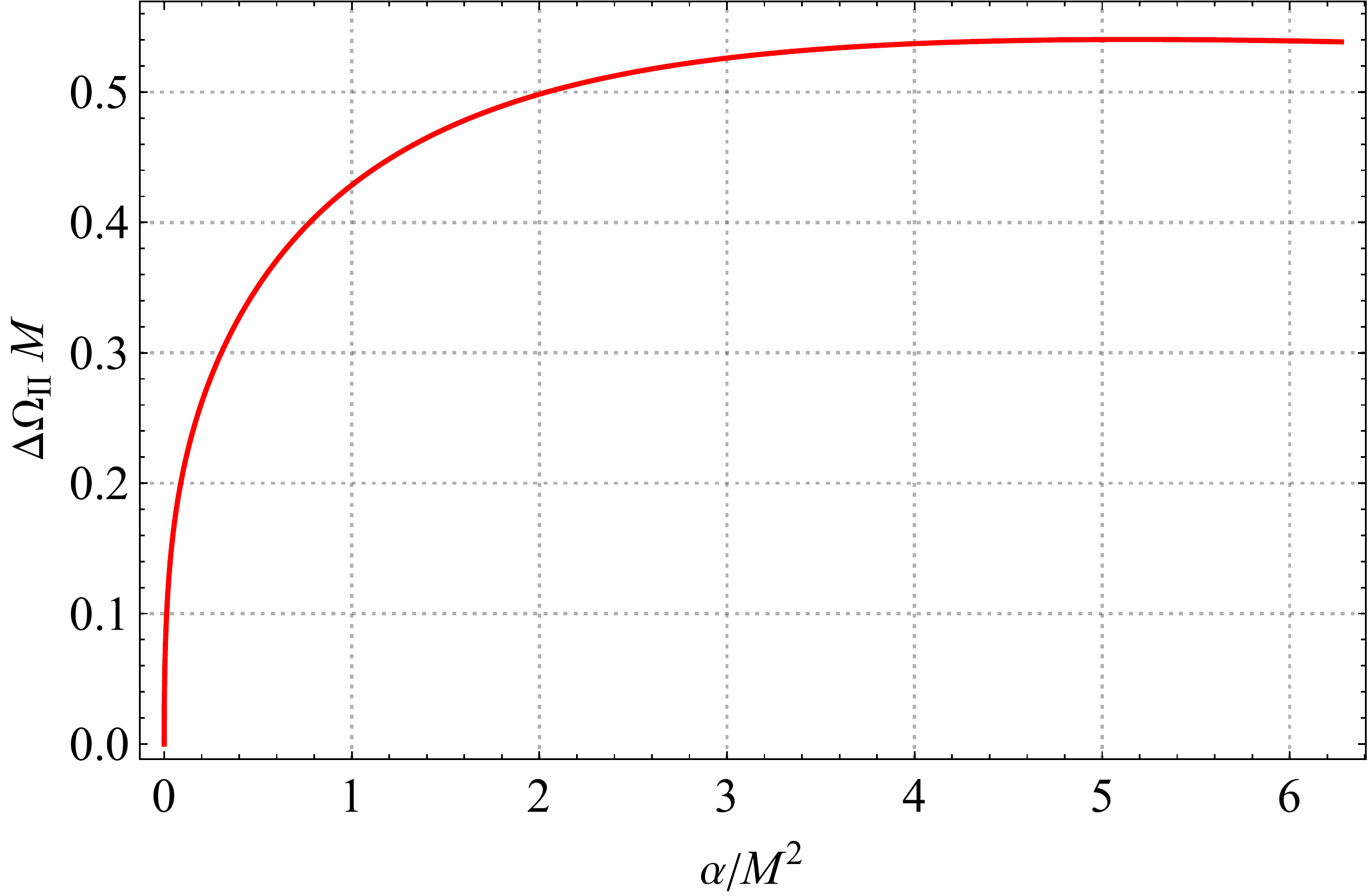}
    \caption{$\partial_M \Delta_\text{equator}(r_\text{min}, M, J)$ and $\Delta \Omega_\text{II}$ of the extremal black hole solution as a function of the coupling constant in the parameter-region II. }
    \label{figII}
\end{figure*}

\subsection{Type-II violation}

Let's proceed to analyze the situation when the coupling constant satisfies $0<\alpha/M^2< 6.2754$, which corresponds to the parameter-region II. In this region, the boundary curve of the black hole solution represents extremal black hole where the event horizon and the inner horizon overlap at the poles $\theta=0, \pi$. To illustrate the mechanism for destroying the event horizon and exposing the singularity, in Fig. \ref{fig3}, we present the profiles of the inner horizon, outer horizon, and singularity in $(\theta, r/M)$ coordinates with $\alpha/M^2=2$ and different angular momenta near the boundary line. From this figure, we observe that by increasing the angular momentum, the inner and event horizons tend to overlap at the poles, and eventually, the event horizon is destroyed when the angular momentum exceeds its extremal value $a=a_\text{ex}$. This scenario is referred to as a type-II violation of the black hole in our study. In this case, whether the solutions describe a black hole or a naked singularity is determined by the judging function $\Delta_\text{poles}^\text{min}$, which represents the minimal value of $\Delta_\text{poles}(r, M, J)$.

When $\Delta_\text{poles}^\text{min} > 0$, the spacetime solution represents a naked singularity, otherwise, it is a black hole. Therefore, the destruction of the black hole after dropping a test particle in parameter-region II requires
\begin{equation}\label{descd2}
\Delta'{}_\text{poles}^\text{min} = \Delta_\text{poles}(\tilde{r}'_{\text{min}}, M+\delta M, J+\delta J)>0,
\end{equation}
where $\tilde{r}'_\text{min}$ is the minimal point of the function $\Delta_\text{poles}(r, M+\delta M, J+\delta J)$. Let $\tilde{r}_\text{min}$ denote the minimal point of $\Delta_\text{poles}(r, M, J)$. Similar to the previous subsection, we assume that the initial state of the spacetime is an extremal black hole, i.e., the solution lies on the blue boundary curve in Fig. \ref{fig1}. Therefore, we have $\tilde{r}_\text{min}=r_H(0)$, which implies
\begin{equation}
\partial_r\Delta_\text{poles}(\tilde{r}_\text{min}, M, J)=\Delta_\text{poles}(\tilde{r}_\text{min}, M, J)=0.
\end{equation}
After dropping a test particle, the minimal point shifts infinitesimally to $\tilde{r}'_\text{min}=\tilde{r}_\text{min}+\delta \tilde{r}_\text{min}$. Utilizing the above results, the destruction condition (\ref{descd2}) implies
\begin{equation}\begin{aligned}\label{descdII}
&\Delta_\text{poles}(\tilde{r}_\text{min}+\delta \tilde{r}_\text{min}, M+\delta M, J+\delta J) \\
&= \frac{\partial \Delta_\text{poles}}{\partial M}\delta M+\frac{\partial \Delta_\text{poles}}{\partial J}\delta J>0,
\end{aligned}\end{equation}
under the first-order approximation of the particle perturbation. In the left panel of Fig. \ref{figII}, we demonstrate that
\begin{equation}
{\partial_r}\Delta_\text{poles}(\tilde{r}_\text{min}, M, J) < 0
\end{equation}
for the extremal black hole solutions in parameter-region II, where $0<\alpha/M^2< 6.2754$. Consequently, the destruction condition (\ref{descdII}) gives
\begin{equation}
\delta M < \Omega_\text{II} \delta J,
\end{equation}
where we define
\begin{equation}
\Omega_\text{II}=-\frac{\partial_J \Delta_\text{poles}(\tilde{r}_\text{min}, M, J)}{\partial_M \Delta_\text{poles}(\tilde{r}_\text{min}, M, J)}.
\end{equation}
This provides an upper bound for $E/L$. Together with the condition (\ref{lowbound}) that the particle can be dropped into the black hole on the equatorial plane, the allowed range of $E/L$ for destroying the black hole is
\begin{equation}
\Omega_H^\text{equator}<E/L<\Omega_\text{II}.
\end{equation}
In the right panel of Fig. \ref{figII}, we illustrate the allowed length
\begin{equation}
\Delta\Omega_\text{II} \equiv \Omega_\text{II}-\Omega_H^\text{equator}
\end{equation}
of $E/L$ as a function of the coupling constant $\alpha/M^2$ in parameter-region II. As a result, we observe that the allowed length $\Delta\Omega$ is positive in the parameter region $0 < \alpha/M^2< 6.2754$, indicating that an extremal rotating black hole in these cases can be overspun by throwing a test particle.

\begin{figure*}
    \centering
    \includegraphics[width=0.371\textwidth]{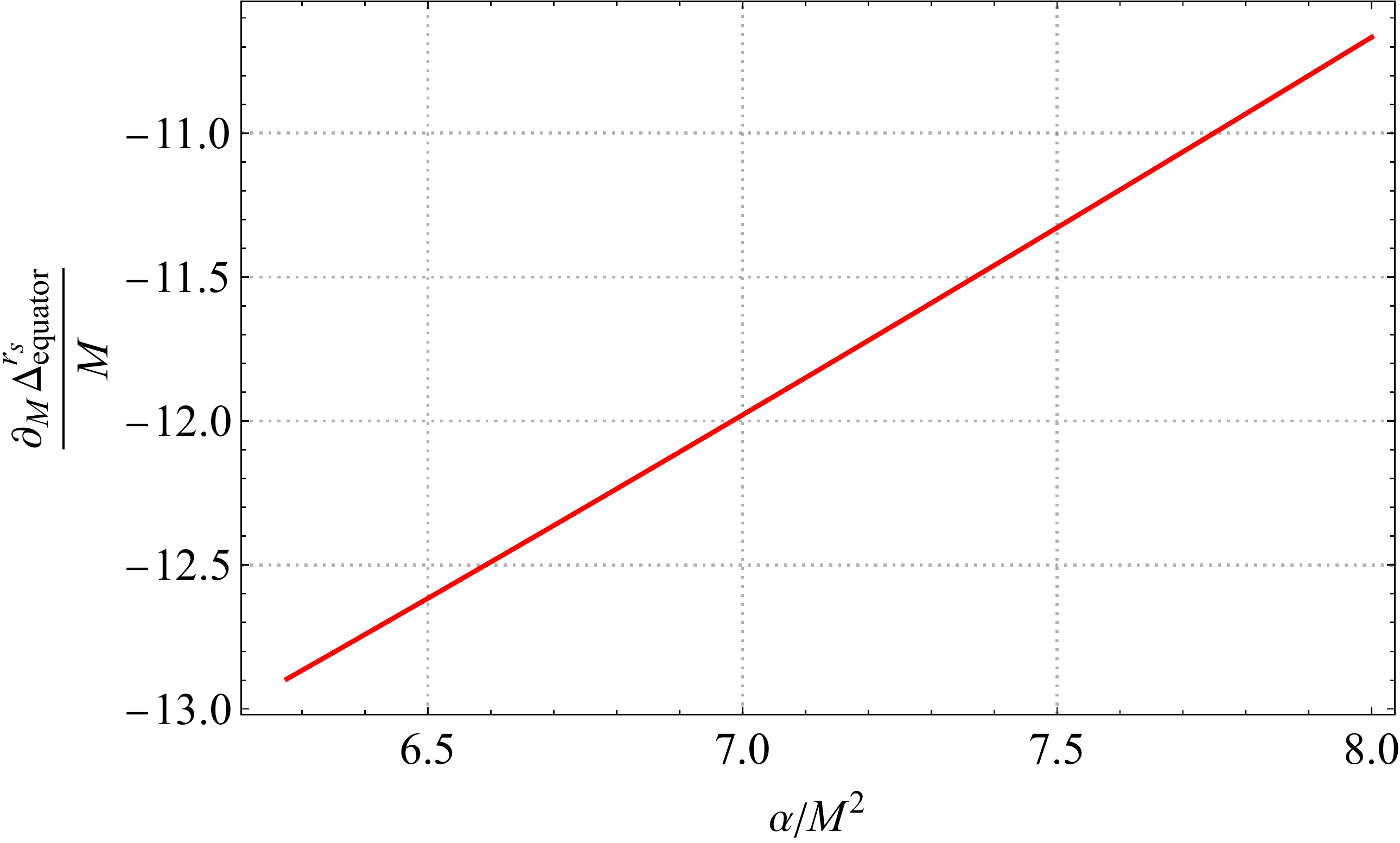}
    \includegraphics[width=0.35\textwidth]{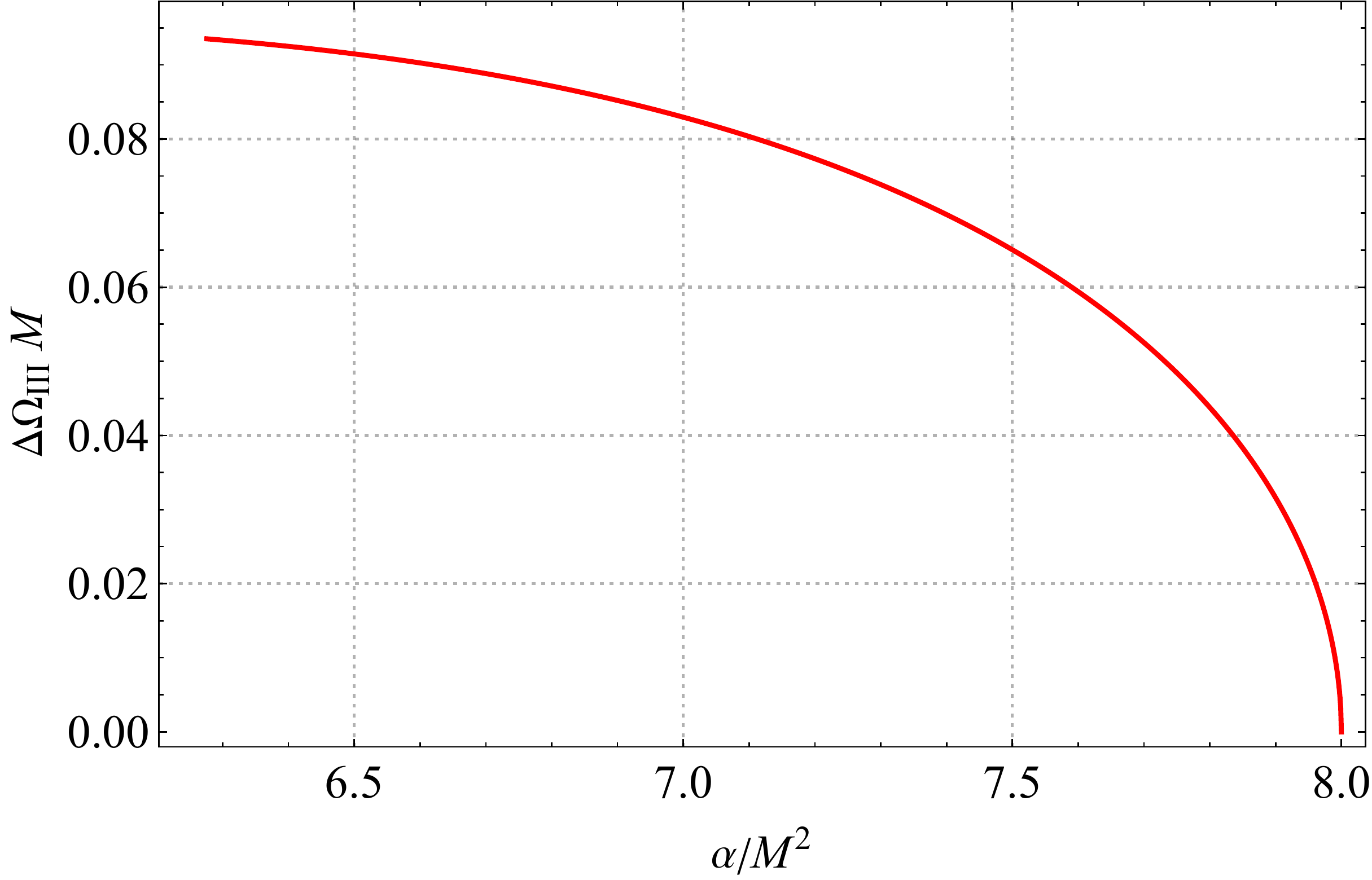}
    \caption{$\partial_M \Delta_\text{equator}(r_\text{min}, M, J)$ and $\Delta \Omega_\text{III}$ of the extremal black hole solution as a function of the coupling constant in the parameter-region III. }
    \label{figIII}
\end{figure*}

\subsection{Type-III violation}

Lastly, let's examine the scenario where the coupling constant satisfies $6.2754 \leq \alpha/M^2 \leq 8$, which corresponds to parameter-region III. In this region, the boundary curve of the black hole solutions represents singular black holes where the event horizon and the singularity overlap at the equator $\theta = \pi/2$. To illustrate the mechanism for destroying the event horizon and exposing the singularity,  in Fig. \ref{fig4}, we present the profiles of the outer horizon and singularity in $(\theta, r/M)$ coordinates with $\alpha/M^2=7$ and different angular momenta near the boundary line. From this figure, we can observe that by increasing the angular momentum, the event horizon and the singularity will tend to overlap at the equator, and eventually, the event horizon is destroyed when the angular momentum exceeds the critical value $a=a_\text{sb}$. This phenomenon is referred to as a type-III violation of the black hole in our paper. To determine whether the solutions describe a black hole or a naked singularity in parameter-region III, we introduce the judging function
\ba\begin{aligned}
\Delta_\text{equator}^{r_s}(J, M)&=\Delta_\text{equator}(r_s)\\
&={J^2}/{M^2}-8 M (\alpha  M)^{1/3}+4 (\alpha  M)^{2/3}\,.
\end{aligned}\ea
When $\Delta_\text{equator}^{r_s} > 0$, the solution corresponds to a naked singularity, otherwise, it describes a black hole. Therefore, the destruction of the black hole after dropping a test particle in parameter-region III requires that
\begin{equation}\label{descd3}
\Delta_\text{equator}^{r_s}(M+\delta M, J+\delta J)>0.
\end{equation}
Since we are considering only the first-order approximation under the perturbation of the test particle, we focus on the scenario where the initial state of the spacetime is a singular black hole, meaning the solution lies on the green boundary curve in Fig. \ref{fig1}, and we have $r_H(\pi/2) = r_s$. Consequently, the destruction condition (\ref{descd3}) can be expressed as
\begin{equation}
\frac{\partial \Delta_\text{equator}^{r_s}}{\partial M}\delta M+\frac{\partial \Delta_\text{equator}^{r_s}}{\partial J}\delta J>0,
\end{equation}
within the first-order approximation of the particle perturbation. In the left panel of Fig. \ref{figIII}, we demonstrate that
\begin{equation}\label{dscdIII}
\partial_M\Delta_\text{equator}^{r_s}( M, J) < 0,
\end{equation}
for the singular black hole solutions in parameter-region III, i.e., $6.2754\leq\alpha/M^2\leq 8$. Based on Eq. (\ref{dscdIII}), we obtain the destruction condition
\begin{equation}
\delta M < \Omega_\text{III} \delta J,
\end{equation}
where we define
\begin{equation}
\Omega_\text{III}=-\frac{\partial_J \Delta_\text{equator}^{r_s}(M, J)}{\partial_M \Delta_\text{equator}^{r_s}(M, J)}.
\end{equation}
This provides an upper bound for $E/L$. Combining it with the condition (\ref{lowbound}) that the particle can be dropped into the black hole on the equatorial plane, we conclude that the allowed range of $E/L$ for destroying the black hole is
\begin{equation}
\Omega_H^\text{equator}<E/L<\Omega_\text{III}.
\end{equation}
In the right panel of Fig. \ref{figIII}, we display the allowed length
\begin{equation}
\Delta\Omega_\text{III} \equiv \Omega_\text{III}-\Omega_H^\text{equator}
\end{equation}
of $E/L$ as a function of the coupling constant $\alpha/M^2$ in parameter-region III. Consequently, we observe that the allowed range $\Delta\Omega_\text{III}$ is positive for $6.2754 \leq \alpha/M^2 < 8$, while $\Delta\Omega_\text{III} = 0$ in the case of the static singular black hole where $\alpha/M^2 = 8$ and $a = 0$. This indicates that, except for the static limit scenario at $\alpha/M^2=8$, the singular black hole can be overspun by introducing a test particle, violating the WCCC.

\section{Conclusion}\label{sec4}

In the present work, we have delved deeply into the WCCC for rotating black holes in semiclassical gravity with a type-A trace anomaly. Specifically, we have meticulously examined how a test particle alters the spacetime of extremal or singular rotating black holes in such effective gravitational theories, unveiling any potential violations of the WCCC. Through our investigations, we found that depending on the three different ranges of the coupling constant $\alpha/M^2$, we can observe three distinct boundary characteristics of the black hole solutions, corresponding to three distinct types of WCCC violations. The three types of violations of WCCC was explored by launching a test particle towards the black hole on the equatorial plane. With the exception of extremal Kerr, static extremal, and static singular black holes (i.e., $\alpha/M^2=0,-1,8$), our results demonstrate that all extremal black holes and singular black holes can be destroyed under the first-order perturbation of the test particle, leading to a violation of the WCCC.

For the non-static cases, the aforementioned analyses have clearly demonstrated the possibility of breaching the WCCC. Taking the Type-I violation as an example, even in near-extremal black hole scenarios, the destruction conditions can still be satisfied as long as the initial state is very close to extremal. Thus, if we can observe that the event horizon of extremal (singular) black holes can be destroyed, then the possibility of destruction in the near-extremal (singular) black holes also exists. Therefore, we would only need to consider the near-extremal (singular) and the second-order perturbation approximations in cases where the black hole cannot be destroyed under the first-order perturbation, which precisely corresponds to the $\alpha/M^2=-1,0,8$ cases in our model. This will be left for future investigations. Furthermore, the equations of particle motion are governed by the classical geodesic equation, with no considerations for the impact of quantum corrections. As such, these findings suggest that the preservation of the WCCC may necessitate the consideration of modifications to the test particle behavior in the presence of quantum effects.

\section*{Acknowledgements}
JJ  is supported by the National Natural Science Founda- tion of China with Grant No. 12205014, the Guangdong Basic and Applied Research Foundation with Grant No. 217200003 and the Talents Introduction Foundation of Beijing Normal University with Grant No. 310432102. MZ is supported by the National Natural Science Foundation of China with Grant No. 12005080.

\end{document}